\documentstyle[12pt,dina4,epsf]{article}
\newcounter{saveeqn}
\renewcommand{\theequation}{\arabic{section}.\arabic{equation}}
\newcommand{\alpheqn}{\setcounter{saveeqn}{\value{equation}}%
\addtocounter{saveeqn}{1}\setcounter{equation}{0}%
\renewcommand{\theequation}{%
 \mbox{\arabic{section}.\arabic{saveeqn}\alph{equation}}}}
\newcommand{\reseteqn}{\setcounter{equation}{\value{saveeqn}}%
\renewcommand{\theequation}{\arabic{section}.\arabic{equation}}}
\newcommand{\appeqn}{%
\renewcommand{\theequation}{\Alph{section}.\arabic{equation}}}
\newcommand{\appfig}{\setcounter{figure}{0}%
\renewcommand{\thefigure}{\Alph{section}.\arabic{figure}}}

\newcommand{\shmg}{\frac{1}{\bh}\sum_{n=-\bh/2}^{\bh/2-1}}
\newcommand{\shmmg}{\frac{1}{\bh}\sum_{m=-\bh/2}^{\bh/2-1}}

\newcommand{\gam}{\gamma_{\mu}}

\newcommand{\kt}{\tilde{k}}
\newcommand{\qt}{\tilde{q}}
\newcommand{\rt}{\tilde{r}}
\newcommand{\st}{\tilde{s}}

\newcommand{\bh}{\hat{\beta}}
\newcommand{\muh}{\hat{\mu}}

\newcommand{\wh}{\hat{\omega}}
\newcommand{\whp}{\hat{\omega}_{n}^{+}}
\newcommand{\whm}{\hat{\omega}_{m}^{-}}

\newcommand{\zli}{\bar{z}_{i}}

\newcommand{\vp}{\vec{p}}
\newcommand{\vk}{\vec{k}}
\newcommand{\vq}{\vec{q}}
\newcommand{\vr}{\vec{r}}
\newcommand{\vs}{\vec{s}}

\newcommand{\im}{\mbox{Im}}

\newcommand{\res}{\mbox{Res}}
\newcommand{\arcosh}{\mbox{arcosh}}

\input{epsf.sty}
\def\axowidth{0.5 }
\def\axoscale{1.0 }
\def\axoxoff{0 }
\def\axoyoff{0 }

\def\Gluon(#1,#2)(#3,#4)#5#6{
\put(\axoxoff,\axoyoff){
}
}

\def\Photon(#1,#2)(#3,#4)#5#6{
\put(\axoxoff,\axoyoff){
}
}

\def\ZigZag(#1,#2)(#3,#4)#5#6{
\put(\axoxoff,\axoyoff){
}
}

\def\PhotonArc(#1,#2)(#3,#4,#5)#6#7{
\put(\axoxoff,\axoyoff){
}
}

\def\GlueArc(#1,#2)(#3,#4,#5)#6#7{
\put(\axoxoff,\axoyoff){
}
}

\def\ArrowArc(#1,#2)(#3,#4,#5){
\put(\axoxoff,\axoyoff){
}
}

\def\LongArrowArc(#1,#2)(#3,#4,#5){
\put(\axoxoff,\axoyoff){
}
}

\def\DashArrowArc(#1,#2)(#3,#4,#5)#6{
\put(\axoxoff,\axoyoff){
}
}

\def\ArrowArcn(#1,#2)(#3,#4,#5){
\put(\axoxoff,\axoyoff){
}
}

\def\LongArrowArcn(#1,#2)(#3,#4,#5){
\put(\axoxoff,\axoyoff){
}
}

\def\DashArrowArcn(#1,#2)(#3,#4,#5)#6{
\put(\axoxoff,\axoyoff){
}
}

\def\ArrowLine(#1,#2)(#3,#4){
\put(\axoxoff,\axoyoff){
}
}

\def\LongArrow(#1,#2)(#3,#4){
\put(\axoxoff,\axoyoff){
}
}

\def\DashArrowLine(#1,#2)(#3,#4)#5{
\put(\axoxoff,\axoyoff){
}
}

\def\Line(#1,#2)(#3,#4){
\put(\axoxoff,\axoyoff){
}
}

\def\DashLine(#1,#2)(#3,#4)#5{
\put(\axoxoff,\axoyoff){
}
}

\def\CArc(#1,#2)(#3,#4,#5){
\put(\axoxoff,\axoyoff){
}
}

\def\DashCArc(#1,#2)(#3,#4,#5)#6{
\put(\axoxoff,\axoyoff){
}
}

\def\Vertex(#1,#2)#3{
\put(\axoxoff,\axoyoff){
}
}

\def\Text(#1,#2)[#3]#4{
\dimen0=\axoxoff \unitlength
\dimen1=\axoyoff \unitlength
\advance\dimen0 by #1 \unitlength
\advance\dimen1 by #2 \unitlength
\makeatletter
\@killglue\raise\dimen1\hbox to\z@{\kern\dimen0 \makebox(0,0)[#3]{#4}\hss}
\ignorespaces
\makeatother
}

\def\BCirc(#1,#2)#3{
\put(\axoxoff,\axoyoff){
}
}

\def\GCirc(#1,#2)#3#4{
\put(\axoxoff,\axoyoff){
}
}

\def\EBox(#1,#2)(#3,#4){
\put(\axoxoff,\axoyoff){
}
}

\def\BBox(#1,#2)(#3,#4){
\put(\axoxoff,\axoyoff){
}
}

\def\GBox(#1,#2)(#3,#4)#5{
\put(\axoxoff,\axoyoff){
}
}

\def\Boxc(#1,#2)(#3,#4){
\put(\axoxoff,\axoyoff){
}
}

\def\BBoxc(#1,#2)(#3,#4){
\put(\axoxoff,\axoyoff){
}
}

\def\GBoxc(#1,#2)(#3,#4)#5{
\put(\axoxoff,\axoyoff){
}
}

\def\SetOffset(#1,#2){\def\axoxoff{#1 } \def\axoyoff{#2 }}

\def\fsize{10 }

\def\PText(#1,#2)(#3)[#4]#5{
\ifx#4 lt{\def\fmode{0 }}\else{
\ifx#4 tl{\def\fmode{0 }}\else{
\ifx#4 lb{\def\fmode{2 }}\else{
\ifx#4 bl{\def\fmode{2 }}\else{
\ifx#4 l{\def\fmode{1 }}\else{
\ifx#4 rt{\def\fmode{6 }}\else{
\ifx#4 tr{\def\fmode{6 }}\else{
\ifx#4 rb{\def\fmode{8 }}\else{
\ifx#4 br{\def\fmode{8 }}\else{
\ifx#4 r{\def\fmode{7 }}\else{
\ifx#4 t{\def\fmode{3 }}\else{
\ifx#4 b{\def\fmode{5 }}\else{ \def\fmode{4 } }\fi
}\fi}\fi}\fi}\fi}\fi}\fi}\fi}\fi}\fi}\fi}\fi
\put(#1,#2){\makebox(0,0)[]{\special{"/pfont findfont /fsize scale setfont
\axoxoff \axoyoff #3 \fmode \fsize (#5) ptext }}}  }

\def\GOval(#1,#2)(#3,#4)(#5)#6{
\put(\axoxoff,\axoyoff){
}
}

\def\Oval(#1,#2)(#3,#4)(#5){
\put(\axoxoff,\axoyoff){
}
}

\let\eind=]

\def\kromme(#1,#2)#3{#1 #2 \ifx #3\eind\else\expandafter\kromme\fi#3}

\def\LogAxis(#1,#2)(#3,#4)(#5,#6,#7,#8){
\put(\axoxoff,\axoyoff){
}
}

\def\LinAxis(#1,#2)(#3,#4)(#5,#6,#7,#8,#9){
\put(\axoxoff,\axoyoff){
}
}

\title{ \begin{large} LATTICE ARTEFACTS IN THE NON-ABELIAN DEBYE \\
 SCREENING MASS IN ONE LOOP ORDER \end{large}  }
\author{ P.Kaste and H.J. Rothe 
    \\ 
 \parbox{\textwidth}{
  \begin{center}
   Institut f\"{u}r Theoretische Physik \\ Universit\"{a}t Heidelberg \\ 
  Philosophenweg 16 \\ D-69120 Heidelberg
 \end{center}
  } }

\begin{document}
\maketitle
\begin{abstract}
We compute the electric screening mass in lattice QCD 
with Wilson fermions at finite temperature
and chemical potential to one-loop order, and 
show that lattice artefacts arising from a finite lattice spacing result
in an enhancement of the screening mass as compared to the continuum.
We discuss the magnitude of this enhancement as a function of the temperature
and chemical potential for lattices with different number of lattice
sites in the temporal direction that can be implemented in lattice 
simulations. Most of the enhancement is found to be due to the fermion
loop contribution.
\end{abstract}

\pagebreak
\section{Introduction}
An important feature of finite temperature QCD is the generation of an
electric and magnetic screening mass which play an important role in
controlling
the infrared behaviour of the theory. The electric screening mass leads to
a Debye
screened static quark-antiquark potential and is given for SU(N) 
with $N_{f}$ quark flavours, and for vanishing chemical potential and 
quark mass, in
leading order
perturbation theory by $m^{2}_{el} = \frac{g^{2}}{3}(N+N_{f}/2)T^{2}$ 
\cite{kala}.
Renormalization group improved perturbation
theory
tells us that the effective coupling is a function of the temperature, and
decreases
with increasing temperature. This suggests that at sufficiently high
temperatures,
above the deconfining phase transition, the screening mass may be computed in
perturbation theory. Because of the singular
infrared behaviour of the perturbative series, however, the computation
of the next to leading order contribution requires a resummation of infrared
divergent diagrams which turns out to be sensitive to the magnetic
screening mass \cite{rebh}.
This mass vanishes in lowest order perturbation theory and is expected to be
of ${\cal O}(g^{2}T)$. The coefficient multiplying $g^{2}T$ turns out however
to be incalculable \cite{lind}.
Making use
of an improved perturbation theory proposed by Braaten and Pisarski 
\cite{braa}, which
resums hard thermal loops, and of a gauge invariant
definition of the
electric screening mass \cite{kobe}, Rebhan 
has calculated the ${\cal O}(g^{3}T^{2})$
corrections to the non-abelian screening mass-squared and has shown that
next to leading order contributions give rise to an enhancement \cite{rebh}.

The lattice formulation of QCD allows one to
determine the electric screening mass non-perturbatively.
The
screening mass is extracted from correlators of Polyakov loops
[6--9], or
from the long distance behaviour of the gluon propagator 
\cite{hell}. For small
quark-antiquark
separations lattice perturbation theory for the Polyakov loop correlation
function
is expected to describe the
Monte Carlo data, since
for a finite lattice volume one is not confronted with the infrared
problems encountered in thermal perturbation theory. This has been checked
in \cite{pete} for the SU(3) gauge theory
by taking careful account of finite size effects and, in particular,
of the zero momentum modes which do not allow
one to take the thermodynamical limit for fixed coupling, as one would do in
standard perturbation theory. 
For larger quark-antiquark separations, beyond the ``perturbative horizon'',
the colour averaged potential is expected to have a Debye screened form.
Monte Carlo simulations confirm this screening picture [6--9].
In the case of pure SU(2) and SU(3) gauge theories the electric screening
mass, when determined from Polyakov loop correlation functions, is found to
be about $10\%$ larger \cite{gao,irba} 
than the leading order perturbative result 
if the temperature dependent coupling constant is 
determined from Polyakov loop correlators in the perturbative region.
As was pointed out by Rebhan, \cite{rebh}, such an enhancement could also
be expected if next to leading order corrections to the continuum screening
mass are taken into account through resummed perturbation theory. A
quantitative comparison with the results obtained in the above simulations
is however very difficult and has, to our knowledge, not been carried out
so far. In contrast to the work of ref.\ \cite{gao,irba}, the electric
screening mass as extracted in ref.\ \cite{hell} from the gluon propagator
in the Landau gauge was found to deviate strongly form the leading order
perturbative result. 

In comparing the Monte Carlo data for the electric screening mass with
leading order, or resummed, perturbation theory it is important to have
an estimate of the size of lattice artefacts to be expected from a finite
lattice spacing.
To obtain such an estimate
we compute the electric screening
mass in one-loop order on the lattice, at finite temperature and
chemical potential, in the infinite volume
limit, and compare it with the continuum. 
For the case of QED with {\it naive} fermions the screening mass has been 
calculated by Pietig \cite{piet}.
The screening mass is calculated
from the zero-momentum
limit of the 44-component of the vacuum polarization tensor
evaluated for vanishing Matsubara
frequency.
In one-loop order this definition of the screening mass is gauge invariant,
and consistent
with the more general gauge invariant definition given in ref.\ \cite{kobe},
where
the screening mass is determined from the position of the
pole
of the gluon propagator for vanishing Matsubara frequency.

The paper is organized as follows: In the following section we calculate the
electric screening mass for QCD in one-loop order for
the case of Wilson fermions. The Feynman
rules and frequency summation formulae required for the computation are
relegated
to two Appendices. As we shall see, the resulting integral expression has a
very transparent form. In section 3
we then evaluate the momentum integrals for the
screening mass numerically and compare the results with the
continuum. We show that the lattice artefacts due to a finite
lattice spacing give rise
to an enhancement of the screening mass as compared to the continuum. We
discuss the magnitude of this enhancement as a function of 
the temperature and chemical potential
for lattices with different number of lattice sites in the
temporal direction which can be implemented in numerical simulations.
Most of the enhancement
is found to be due to the fermion
loop contribution.
Section
4 contains a summary of our results.
\section{Electric Screening Mass in One Loop Order}
In this section we compute the electric screening mass in lattice QCD to
one loop order
from the zero momentum limit of the 44-component of the vacuum polarization
tensor evaluated for vanishing Matsubara frequency. The Feynman diagrams
contributing
in this order are shown in fig.\ 1. While diagrams (a,c,d,e) have a
continuum analog, the
remaining diagrams, required by gauge invariance on the lattice, do not
possess a
continuum counterpart.
\begin{figure}[ht]
\leavevmode
\centering
\epsfxsize10cm \epsffile{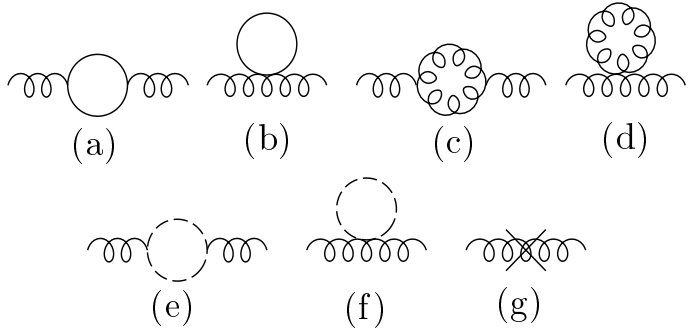}
\caption{Feynman diagrams contributing to the vacuum polarization in
  one loop order on the lattice.}
\end{figure}
The finite temperature, finite chemical potential
lattice Feynman
rules are obtained from the $T=\mu =0$ rules by replacing the fourth
component of
the fermion and boson momenta by $\hat{\omega}^{-}_{\ell} +i\hat{\mu}$ and
$\hat{\omega}^{+}_{\ell}$,
respectively, where  $\hat{\omega}^{+}_{\ell} = \frac{2\pi}{\hat{\beta}}\ell
\ (\hat{\omega}^{-}_{\ell} = \frac{(2\ell+1)\pi}{\hat{\beta}})$, with
$\ell \in Z$, 
are the Matsubara
frequencies
for bosons (fermions), and $\hat{\beta}$ is the inverse temperature.
Quantities with a
"hat" are always understood to be measured in lattice units. Furthermore,
integrals over the
fourth component of momenta at zero temperature are replaced at finite
temperature by
sums over Matsubara frequencies in the interval $[-\frac{\hat{\beta}}{2},
\frac{\hat{\beta}}{2}-1]$, 
where we have taken $\hat{\beta}$ to be even. The
Feynman rules
are collected in Appendix A. The relevant formulae for carrying out the sums
over Matsubara frequencies are derived in Appendix B.

The contributions of diagrams (a-g) to the vacuum polarization tensor are
diagonal
in colour space,
\begin{equation}
\hat{\Pi}^{(\beta,\mu)AB}_{\mu\nu}(\hat{\omega}^{+}_{\ell},\vec{\hat{k}}) =
\delta_{AB}\hat{\Pi}^{(\beta,\mu)}_{\mu\nu}(\hat{\omega}^{+}_{\ell},\vec{\hat{
k}})\ . \label{diag:col}
\end{equation}
The electric screening mass (in lattice units) is then defined by
\begin{equation}
\hat{m}_{el}^{2} = \lim_{\vec{\hat{k}}\rightarrow 0}
\hat{\Pi}^{(\beta,\mu)}_{44}(0,\vec{\hat{k}})\ . 
\end{equation}
In the following we first consider the contributions to
$\hat{\Pi}^{(\beta,\mu)}_{44}(0,\vec{\hat{k}})$ coming from the fermion
loops, i.e.,
diagrams (a) and (b).\\ \\
{\it i) Contribution of diagram (a)}\\ 

A straight forward application of the finite temperature, finite chemical
potential
lattice Feynman rules yields
\alpheqn
\begin{equation}
\hat{\Pi}^{(\beta,\mu)}_{44}(0,\vec{\hat{k}})_{(a)}
=-\frac{N_{f}}{2}g^{2}\frac{1}{\hat{\beta}}
 \sum^{\frac{\hat{\beta}}{2}-1}_{\ell=-\frac{\hat{\beta}}{
 2}}\int^{\pi}_{-\pi}\frac{d^{3}\hat{p}}{(2\pi)^{3}}f^{(a)}
(e^{i(\hat{\omega}_{\ell}^{-}+i\hat{\mu})};\vec{\hat{p}},
 \vec{\hat{k}})\ , \label{pi:a}
\end{equation}
where
\begin{equation}
f^{(a)}(z;\vec{\hat{p}},\vec{\hat{k}})=
\frac{2(z^{4}+1)-2\eta(z^{3}+z)+4\xi{\cal G}z^{2}}{\Pi^{4}_{i=1}(z-z_i)}
\end{equation}
with
\begin{eqnarray}
\eta &=& \frac{1}{[1+\hat{M}(\vec{\hat{p}})]} + \frac{1}{[1+\hat{M}
 (\vec{\hat{p}}+\vec{\hat{k}})]} \ , \\
\xi &=& \frac{1}{[1+\hat{M}(\vec{\hat{p}})][1+\hat{M}(\vec{\hat{p}}+
 \vec{\hat{k}})]} \ , \\
{\cal G} &=& 1+\sum_{j}\sin\hat{p}_{j}\sin(\hat{p}+\hat{k})_{j} \ , \\
\hat{M}(\vec{\hat{p}}) &=& \hat{m} +2 \sum_{j} \sin^{2}\frac{\hat{p}_{j}}{2}\ ,
\end{eqnarray}
\reseteqn
and $N_{f}$ is the number of quark-flavours.
The position of the poles of $f^{(a)}(z;\vec{\hat{p}},\vec{\hat{k}})$ 
are given by
\alpheqn
\begin{eqnarray}
z_{1} &=& e^{\phi};\ \ \ z_{2}=e^{-\phi} \ , \nonumber \\
z_{3} &=& e^{\psi};\ \ \ z_{4}=e^{-\psi} \ ,
\end{eqnarray}
with
\begin{eqnarray}
\phi &=& \tilde{\cal E}(\vec{\hat{p}}) \ , \nonumber \\
\psi &=& \tilde{\cal E}(\vec{\hat{p}}+\vec{\hat{k}})\ ,
\end{eqnarray}
where
\begin{equation}
\tilde{\cal E}(\vec{\hat{q}}) = \ln\left[ K(\vec{\hat{q}}) +
 \sqrt{K^{2}(\vec{\hat{q}})-1}\, \right]
 = \arcosh K(\vec{\hat{q}})
\end{equation}
and
\begin{equation}
K(\vec{\hat{q}}) = 1 + \frac{\bar{E}^{2}(\vec{\hat{q}})}{2[1+
 \hat{M}(\vec{\hat{q}})]}\ , \ \
\bar{E}(\vec{\hat{q}}) = \sqrt{\sum_{j} \sin^{2} \hat{q}_{j} +
 \hat{M}^{2}(\vec{\hat{q}})}\ .
\end{equation}
\reseteqn
Note that
\begin{equation}
\phi \stackrel{\vec{\hat{p}} \rightarrow
-\vec{\hat{p}}-\vec{\hat{k}}}{\longleftrightarrow} \psi \ , \label{subst}
\end{equation}
while $\eta$, $\xi$ and ${\cal G}$ are invariant under the
transformation $\vec{p} \rightarrow -\vec{p}-\vec{k}$. 
This will be important further
below.
The frequency sum can be performed making use of eq.\ (\ref{sum:fermg}) 
derived in
Appendix B. One
finds that
\alpheqn
\begin{eqnarray}
\frac{1}{\hat{\beta}}\sum^{\frac{\hat{\beta}}{2}-1}_
{\ell=-\frac{\hat{\beta}}{2}}f^{(a)}
 (e^{i(\hat{\omega}_{\ell}^{-}+i\hat{\mu})};
\vec{\hat{p}},\vec{\hat{k}}) &=& 2 + h(\phi,\psi,\eta,\xi,{\cal G})
\left[ \frac{1}{e^{\hat{\beta}(\phi+\hat{\mu})}+1} -
 \frac{1}{e^{-\hat{\beta}(\phi-\hat{\mu})}+1} \right] \nonumber \\
& & \mbox{} + h(\psi,\phi,\eta,\xi,{\cal G})
\left[ \frac{1}{e^{\hat{\beta}(\psi+\hat{\mu})}+1} -
 \frac{1}{e^{-\hat{\beta}(\psi-\hat{\mu})}+1} \right]\ , 
 \nonumber \\ \label{eq:aa}
\end{eqnarray}
where
\begin{equation}
h(\phi,\psi,\eta,\xi,{\cal G})= \frac{\cosh 2\phi-\eta\cosh\phi
+\xi{\cal G}}{\sinh\phi(\cosh\phi-\cosh\psi)}\ . \label{eq:ab}
\end{equation}
\reseteqn
To obtain $\hat{\Pi}^{(\beta,\mu)}_{44}(0;\vec{\hat{k}})_{(a)}$, 
we must integrate
this expression over $\vec{\hat{p}}$, with $\hat{p}_{j}
\in [-\pi,\pi]$. Noting that $\eta$, $\xi$ and
${\cal G}$ are invariant under the transformation $\vec{\hat{p}}
\rightarrow -\vec{\hat{p}}-\vec{\hat{k}}$,
and making use of (\ref{subst}), 
as well as of the fact that the integrand in (\ref{pi:a})
is a periodic function in $\hat{p}_{j}$ and $\hat{k}_{j}$, we
can combine the last two contributions on the r.h.s. of (\ref{eq:aa}) 
and obtain
\alpheqn
\begin{eqnarray}
\hat{\Pi}^{(\beta,\mu)}_{44}(0,\vec{\hat k})_{(a)} &=&
 N_{f}g^{2}\int^{\pi}_{-\pi}\frac{d^3\hat{p}}{(2\pi)^{3}}
 [h(\phi,\psi,\eta,\xi,
 {\cal G})-1] \nonumber \\
& &\mbox{} -N_{f}g^{2}\int^{\pi}_{-\pi}\frac{d^3\hat{p}}{(2\pi)^{3}}
 h(\phi,\psi,
 \eta,\xi,{\cal G})[\hat{\eta}_{FD}(\phi)+\bar{\hat{\eta}}_{FD}
 (\phi)]\ . \label{erg:a}
\end{eqnarray}
where
\begin{equation}
\hat{\eta}_{FD}(\phi)=\frac{1}{e^{\hat{\beta}(\phi-\hat{\mu})}+1}\ ,\ \
\bar{\hat{\eta}}_{FD}(\phi)=\frac{1}{e^{\hat{\beta}
 (\phi+\hat{\mu})}+1}\ ,
\end{equation}
\reseteqn
are the lattice Fermi-Dirac distribution functions
for particles and antiparticles.\\ \\
{\it ii) Contribution of diagram (b)} \\

We next compute the contribution to $\hat{\Pi}^{(\beta,\mu)}_{44}(0,
\vec{\hat{k}})$ of the Feynman diagram (b) depicted in Fig.\ 1. 
This diagram has
no analog in the continuum and is given by
\alpheqn
\begin{equation}
\hat{\Pi}^{(\beta,\mu)}_{44}(0,\vec{\hat{k}})_{(b)} = - N_{f} g^{2}
 \frac{1}{\hat{\beta}}
\sum^{\frac{\hat{\beta}}{2}-1}_{\ell=\frac{\hat{\beta}}{2}}\int^{\pi}_{-\pi}
 \frac{d^{3}\hat{p}}{(2\pi)^{3}} f^{(b)}(e^{i(\hat{\omega}_{\ell}^{-}+i
 \hat{\mu})},\vec{\hat{p}} )
\end{equation}
where
\begin{eqnarray}
 f^{(b)}(z;\vec{\hat{p}}) &=& -\frac{z^2-2\rho z+1}{(z-z_1)(z-z_2)}\ , \\
 \rho &=& \frac{1}{1+\hat{M}(\vec{\hat{p}})}\ ,
\end{eqnarray}
\reseteqn
and where $z_{1}$ and $z_{2}$ have been defined
in (2.4a-d). Making again use of the frequency summation formula
(\ref{sum:fermg}), one verifies that
\begin{eqnarray}
\lefteqn{
 \hat{\Pi}^{(\beta,\mu)}_{44}(0,\vec{\hat{k}})_{(b)}=-N_{f}g^{2}
 \int^{\pi}_{-\pi}
 \frac{d^{3}\hat{p}}{(2\pi)^{3}} \left(\coth \phi- \frac{\rho}{\sinh
 \phi}-1 \right) } \nonumber \\
& & \mbox{} + N_{f} g^{2} \int^{\pi}_{-\pi}
 \frac{d^{3}\hat{p}}{(2\pi)^{3}} \left(\coth \phi- \frac{\rho}{\sinh
 \phi} \right)[\hat{\eta}_{FD}(\phi)
 +\bar{\hat{\eta}}_{FD}(\phi)]\ .
\end{eqnarray}
Combining this expression with (\ref{erg:a}) one finds that
\alpheqn
\begin{equation}
\hat{\Pi}^{(\beta,\mu)}_{44}(0,\vec{\hat{k}})
 =\hat{\Pi}^{(vac)}_{44}(0,\vec{\hat{k}}) +
 N_{f} g^{2} \int^{\pi}_{-\pi}
 \frac{d^{3}\hat{p}}{(2\pi)^{3}}H(\phi,\psi,\rho,\eta,\xi,{\cal G})
 [ \hat{\eta}_{FD}(\phi)+\bar{\hat{\eta}}_{FD}(\phi) ]\ , \label{eq:bc}
\end{equation}
where
\begin{equation}
 H(\phi,\psi,\rho,\eta,\xi,{\cal G}) = \coth \phi - \frac{\rho}{\sinh\phi}-
  h(\phi,\psi,\eta,\xi,{\cal G})
\end{equation}
and
\begin{equation}
 \hat{\Pi}^{(vac)}_{44}(0,\vec{\hat{k}}) = -N_{f}g^{2} \int^{\pi}_{-\pi}
 \frac{d^{3}\hat{p}}{(2\pi)^{3}}H(\phi,\psi,\rho,\eta,\xi,{\cal G}
 ) \label{eq:ba}
\end{equation}
\reseteqn
is the $T = \mu = 0$ contribution.
As we now show, $\hat{\Pi}^{(vac)}_{44}(0,\vec{\hat{k}})$ vanishes in the limit
$\vec{\hat{k}} \rightarrow 0$, and hence does not contribute to
the screening mass.

Consider the function $h(\phi,\psi,\eta,\xi,{\cal G})$ defined in
(\ref{eq:ab}). It is singular for $\vec{\hat{k}} \rightarrow 0$, 
since in this
limit $\psi \rightarrow \phi$. 
The singularity is however integrable.\footnote{We
follow here
and in the following a technique used in ref. \cite{piet}, where the author
has calculated the screening mass for naive
fermions in lattice QED to one loop order}
This can be seen as follows. Since according to (\ref{subst}), 
and the statement
following it
\begin{equation}
 h(\phi,\psi,\eta,\xi,{\cal G}) 
 \stackrel{
 \vec{\hat{p}}\rightarrow -\vec{\hat{p}}-\vec{\hat{k}}}{\longrightarrow}
 h(\psi,\phi,\eta,\xi,{\cal G})
\end{equation}
we can also write (\ref{eq:ba}) in the form
\alpheqn
\begin{eqnarray}
 \hat{\Pi}^{(vac)}_{44}(0,\vec{\hat{k}})
&=& -N_{f}g^{2}\int^{\pi}_{-\pi}
 \frac{d^{3}\hat{p}}{(2\pi)^{3}}\left(\coth\phi-\frac{\rho}{
 \sinh\phi}\right) \nonumber \\
& &\mbox{}+\frac{1}{2}N_{f}g^{2}\int^{\pi}_{-\pi}\frac{d^{3}\hat{p}}{
 (2\pi)^{3}}
 \tilde{h}(\phi,\psi,\eta,\xi,{\cal G}) \ , \label{eq:bb}
\end{eqnarray}
where
\begin{equation}
 \tilde{h}(\phi,\psi,\eta,\xi,{\cal G})=h(\phi,\psi,\eta,\xi,{\cal G})
  +h(\psi,\phi,\eta,\xi,{\cal G}) \ . \label{eq:bd}
\end{equation}
\reseteqn
Although each term on the rhs of (\ref{eq:bd}) is singular for 
$\vec{\hat{k}} \rightarrow 0$
$(\psi \rightarrow \phi)$, the sum possesses a finite
limit.
Thus setting $\psi=\phi+\epsilon$ and taking the limit 
$\vec{\hat{k}} \rightarrow 0$
($\epsilon \rightarrow 0$),
one verifies that
\begin{equation}
 \lim_{\vec{\hat{k}} \rightarrow 0}\tilde{h}(\phi,\psi,\eta,\xi,{\cal G})
 = 2\left( \coth\phi-\frac{\rho}{\sinh\phi} \right) \ . \label{eq:be}
\end{equation}
From (\ref{eq:bb}) we therefore conclude that
\[ \lim_{\vec{\hat k}\to 0}\hat\Pi^{(vac)}_{44}(0,\vec{\hat k})
=0 \ . \]
This result is not unexpected, since for vanishing temperature and chemical
potential it is
well known in the continuum formulation, that Lorentz and gauge invariance
protects the gluon from acquiring a mass. The screening mass is therefore
determined by the
finite temperature (f.T.), finite chemical potential contribution, given by the
integral in (\ref{eq:bc}). By making again use of the fact that
$\phi\leftrightarrow\psi$, when $\vec{\hat{p}} \rightarrow -
\vec{\hat{p}}-\vec{\hat{k}}$, while $\eta$, $\xi$ and ${\cal G}$
remain invariant under this change of variables, we can write this
contribution in
the form
\begin{eqnarray}
 \hat{\Pi}^{(\beta,\mu)}_{44}(0,\vec{\hat{k}})_{f.T.} &=& N_{f}g^{2}
 \int^{\pi}_{-\pi}
 \frac{d^{3}\hat{p}}{(2\pi)^{3}}\left( \coth\phi-\frac{\rho}{
 \sinh\phi} \right) [ \hat{\eta}_{FD}(\phi)+\bar{\hat{\eta}}_{FD}
(\phi) ] \nonumber \\
& &\mbox{} -\frac{1}{2}N_{f}g^{2}\int^{\pi}_{-\pi}
 \frac{d^{3}\hat{p}}{(2\pi)^{3}} \left\{ h(\phi,\psi,\eta,\xi,{\cal G})
[ \hat{\eta}_{FD}(\phi)+\bar{\hat{\eta}}_{FD}(\phi) ]\right. \nonumber \\
& &\mbox{}+ \left. h(\psi,\phi,\eta,\xi,{\cal G})[ \hat{\eta}_{FD}(\psi)
+\bar{\hat{\eta}}_{FD}(\psi)] \right\} \ . \label{pi:fft}
\end{eqnarray}
Consider the second integral. It can be rewritten as follows
\[ \int^{\pi}_{-\pi}
 \frac{d^{3}\hat{p}}{(2\pi)^{3}} \{ \tilde{h}(\phi,\psi,\eta,\xi,{\cal G})
[\hat{\eta}_{FD}(\psi)+\bar{\hat{\eta}}_{FD}(\psi)]+
 h(\phi,\psi,\eta,\xi,{\cal G})
 \bigtriangleup\hat{\eta}_{FD}(\phi,\psi) \} \ , \]
where $\tilde{h}$ has been defined in (\ref{eq:bd}), and
\[ \bigtriangleup\hat{\eta}_{FD}(\phi,\psi) =
 [\hat{\eta}_{FD}(\phi)-\hat{\eta}_{FD}(\psi)]+
 [\bar{\hat{\eta}}_{FD}(\phi)-\bar{\hat{\eta}}_{FD}(\psi)]\ . \]
According to (\ref{eq:be}), $\tilde{h}$ approaches a finite limit for
$\vec{\hat{k}} \rightarrow 0$. 
Upon making the change of variables 
$\vec{\hat{p}} \rightarrow -\vec{\hat{p}}-\vec{\hat{k}}$
the contribution proportional to $\tilde{h}$ is seen to be
cancelled by
the first integral in (\ref{pi:fft}). We therefore conclude that
\begin{equation}
 \lim_{\vec{\hat{k}} \rightarrow 0}\hat{\Pi}^{(\beta,\mu)}_{44}
 (0,\vec{\hat{k}})=
 -\frac{1}{2}N_{f}g^{2}\lim_{\vec{\hat{k}}\rightarrow 0}\int^{\pi}_{-\pi}
 \frac{d^{3}\hat{p}}{(2\pi)^{3}}
 h(\phi,\psi,\eta,\xi,{\cal G})\bigtriangleup\hat{\eta}_{FD}
 (\phi,\psi)\ .
\end{equation}
We have now dropped the subscript "f.T.", since in this
limit only (\ref{pi:fft}) contributes to the screening mass. To calculate
this limit we proceed as before and set $\psi=\phi+\epsilon$.
One then verifies that for $\epsilon \rightarrow 0$ 
(or $\vec{\hat{k}} \rightarrow 0$) the
fermionic
contribution to the screening mass squared is given by
\begin{equation}
 [\hat{m}_{el}^{2}(\hat{\beta},\hat{\mu},\hat{
m})]_{ferm}=N_{f}g^{2}\hat{\beta}\int^{\pi}_{-\pi}
 \frac{d^{3}\hat{p}}{(2\pi)^{3}}
 \left\{ \frac{e^{\hat{\beta}(\phi+\hat{\mu})}}{
 [e^{\hat{\beta}(\phi+\hat{\mu})}+1]^2}+
 \frac{e^{\hat{\beta}(\phi-\hat{\mu})}}{
 [e^{\hat{\beta}(\phi-\hat{\mu})}+1]^2}\right\} \ . \label{mel:fermg}
\end{equation}
In the continuum limit the electric screening mass is given by
\begin{equation}
 m^{2}_{el} = \lim_{a\rightarrow 0}\frac{1}{a^{2}} \hat{m}_{el}^{2} 
  (\frac{\beta}{a},\mu a,ma)\ . \label{mel:lim}
\end{equation}
For Wilson fermions, only momenta $\vec{\hat{p}}$ in the immediate
neighbourhood of
$\vec{\hat{p}} = 0$ contribute to the integral (\ref{mel:fermg}) for 
$\hat{\beta} \rightarrow \infty$,
$\hat{\beta}\hat{\mu} = \beta\mu$, $\hat{\beta}\hat{m} = \beta m$ fixed.
But in this
limit $\hat{\beta}\phi(\vec{\hat{p}}) \rightarrow 
\beta\sqrt{\vec{p}^{\, 2}+m^{2}}$.
Introducing in (\ref{mel:fermg}) the dimensioned
momenta $\vec{p} = \vec{\hat{p}}/a$ as new integration variables, with $a$
the lattice
spacing, one then verifies, after performing
the angular integration, and a partial integration that
\begin{equation}
 [m^{2}_{el}]_{ferm}=N_{f}\frac{g^{2}}{2\pi^{2}}\int^{\infty}_{0}dp
 \frac{2p^{2}+m^{2}}{\sqrt{p^{2}+m^{2}}}\left[\eta_{FD}(E,\mu) +
 \bar{\eta}_{FD}(E,\mu)\right]\ ,
\end{equation}
where $E = \sqrt{\vec{p}^{\, 2}+m^{2}}$, and
\begin{equation}
 \eta_{FD}(E,\mu) = \frac{1}{e^{\beta (E-\mu)}+1}\ ; \ \
 \bar{\eta}_{FD}(E,\mu) = \frac{1}{e^{\beta (E+\mu)}+1} 
\end{equation}
are the Fermi-Dirac distribution functions for particles and antiparticles.

We next consider the contribution to the electric screening mass arising
from diagrams
(c-g) in fig.\ 1.
They only involve sums over Matsubara frequencies
of the bosonic type. Since in the continuum limit the only dimensioned
scale is the temperature, their contribution to the screening mass will be
of the form $const \times g T$. For finite lattice spacing, however, the
temperature dependence will be modified by lattice artefacts. In
the following we first consider diagrams (c-e) which have an analog
in the continuum.\\ \\
{\it iii) Contribution of diagram (c)}\\

Using the lattice Feynman rules given in Appendix A one finds after some
algebra that
this diagram contributes as follows to
$\hat{\Pi}^{(\beta,\mu)}_{44}(0,\vec{\hat{k}})$,
defined in (\ref{diag:col}),
\alpheqn
\begin{equation}
\hat{\Pi}^{(\beta,\mu)}_{44}(0,\vec{\hat{k}})_{(c)}=\frac{3}{2}
 g^{2}\frac{1}{\hat{\beta}}\sum^{\frac{\hat{\beta}}{2}-1}_{\ell=
 -\frac{\hat{\beta}}{2}} \int^{\pi}_{-\pi}\frac{d^{3}\hat{q}}{
 (2\pi)^{3}}f^{(c)}(e^{i\hat{\omega}^{+}_{\ell}};\vec{\hat{q}},
 \vec{\hat{k}})\ ,
\end{equation}
where
\begin{equation}
 f^{(c)}(z;\vec{\hat{q}},\vec{\hat{k}})=\frac{a(\vec{\hat{k}})(z^{2}-1)^{2}-b
 (\vec{\hat{q}},\vec{\hat{k}})z(z+1)^{2}}{
 \Pi^{4}_{i=1}[z-\bar{z}_i]} \label{eq:ca}
\end{equation}
and
\begin{eqnarray}
 a(\vec{\hat{k}}) &=& \sum_{j}\cos^{2}\frac{\hat{k}_{j}}{2}\ , \\
 b(\vec{\hat{q}},\vec{\hat{k}}) &=& \frac{1}{4} \left[ \sum_{j}
 \widetilde{(\hat{q}-\hat{k})}^{2}_{j} +
 \sum_{j}\widetilde{(\hat{q}+2\hat{k})}^{2}_{j} \right] \ .
\end{eqnarray}
\reseteqn
Here $\tilde{\hat{p}}$ is generically defined by
$\tilde{\hat{p}}_{\mu}=2\sin\frac{\hat{p}_{\mu}}{2}$.
The zeros of the denominator in (\ref{eq:ca}) are located at
\alpheqn
\begin{eqnarray}
 \bar{z}_{1} &=& e^{\tilde{\phi}}\ ;\ \bar{z}_{2}=e^{-\tilde{\phi}}\ ,
  \nonumber \\
 \bar{z}_{3} &=& e^{\tilde{\psi}}\ ;\ \bar{z}_{4}=e^{-\tilde{\psi}}\ ,
\end{eqnarray}
where
\begin{eqnarray}
 \tilde{\phi} &=& \arcosh H(\vec{\hat{q}})\ , \nonumber \\
 \tilde{\psi} &=& \arcosh H(\vec{\hat{q}}+\vec{\hat{k}})\ , \\
 H(\vec{\hat{p}}) &=& 1+2\sum_{j}\sin^{2}\frac{\hat{p}_{j}}{2}\ . \nonumber
\end{eqnarray}
\reseteqn
The frequency sum can be calculated by making use of 
(\ref{sum:bosg}). After some
straight forward algebra one finds that
\alpheqn
\begin{eqnarray}
\lefteqn{
 \hat{\Pi}^{(\beta,\mu)}_{44}(0,\vec{\hat{k}})_{(c)}= 6g^2
 \int^{\pi}_{-\pi}\frac{d^{3}\hat{q}}{(2\pi)^{3}}
 h(\tilde{\phi},\tilde{\psi},
 a,b)\hat{\eta}_{BE}(\tilde{\phi}) } \nonumber \\
& &\mbox{}+\frac{3}{2} g^{2} \left\{ a(\vec{\hat{k}})+\int^{\pi}_{-\pi}
 \frac{d^{3}\hat{q}}{(2\pi)^{3}}[h(\tilde{\phi},\tilde{\psi},a,b)+h
 (\tilde{\psi},\tilde{\phi},a,b)] \right\}\ , \label{eq:cb}
\end{eqnarray}
where
\begin{equation}
 h(\tilde{\phi},\tilde{\psi},a,b)=\frac{-a\, \sinh^{2}\tilde{\phi}
 +\frac{1}{2} b[\cosh\tilde{\phi}+1]}{\sinh\tilde{\phi}
 [\cosh\tilde{\phi}-\cosh\tilde{\psi}]} \label{eq:cc}
\end{equation}
and
\begin{equation}
 \hat{\eta}_{BE}(\tilde{\phi}) = \frac{1}{e^{\hat{\beta}\tilde{\phi}}-1}
\end{equation}
\reseteqn
is the lattice version of the Bose-Einstein distribution function.

In obtaining this result we have made use of the fact that
\begin{equation}
 \tilde{\phi} \stackrel{\vec{\hat{q}} \rightarrow
 -\vec{\hat{q}}-\vec{\hat{k}}}{\longleftrightarrow} \tilde{\psi} \label{substb}
\end{equation}
while $a(\vec{\hat{k}})$ and $b(\vec{\hat{q}},\vec{\hat{k}})$ are invariant
under the
transformation 
$\vec{\hat{q}} \rightarrow -\vec{\hat{q}}-\vec{\hat{k}}$. Note that the
function
(\ref{eq:cc}) is singular for $\vec{\hat{k}} \rightarrow 0$,
since in this limit $\tilde{\phi} \rightarrow \tilde{\psi}$. 
The singularity is however
integrable as can be seen by making use of (\ref{substb}) to write
(\ref{eq:cb}) in the form
\alpheqn
\begin{eqnarray}
\lefteqn{
 \hat{\Pi}^{(\beta,\mu)}_{44}(0,\vec{\hat{k}})_{(c)}=3g^2
 \int^{\pi}_{-\pi}\frac{d^{3}\hat{q}}{(2\pi)^{3}}
 h(\tilde{\phi},\tilde{\psi},a,b)
 \bigtriangleup\hat{\eta}_{BE}(\tilde{\phi},\tilde{\psi}) } \\
& &\mbox{}+\frac{3}{2}g^{2} \left\{ a(\vec{\hat{k}})+\int^{\pi}_{-\pi}
 \frac{d^{3}\hat{q}}{(2\pi)^{3}}[h(\tilde{\phi},\tilde{\psi},a,b)+
 h(\tilde{\psi},
 \tilde{\phi},a,b)][1+2\hat{\eta}_{BE}(\tilde{\phi})] \right\}\ , \nonumber
\end{eqnarray}
where
\begin{equation}
 \bigtriangleup\hat{\eta}_{BE}(\tilde{\phi},\tilde{\psi}) = 
 \hat{\eta}_{BE}(\tilde{\phi})
 -\hat{\eta}_{BE}(\tilde{\psi}) \ . \label{def:delbe}
\end{equation}
\reseteqn
The limit $\vec{\hat{k}} \rightarrow 0$ 
can now be easily taken and one obtains the
following contribution to the electric screening mass
\begin{eqnarray}
 (\hat{m}^{2}_{el})_{(c)} &=& \frac{3}{2} g^{2} \left\{ 3-\int^{\pi}_{-\pi}
 \frac{d^{3}\hat{q}}{(2\pi)^{3}} \left[
 3\coth\tilde{\phi}+\frac{1}{2\sinh
 \tilde{\phi}} \right] [1+2\hat{\eta}_{BE}(\tilde{\phi})] \right\} \nonumber \\
& &\mbox{}+\frac{15}{2} g^{2}\hat{\beta}\int^{\pi}_{-\pi}
 \frac{d^{3}\hat{q}}{(2\pi)^{3}}
 \frac{e^{\hat{\beta}\tilde{\phi}}}{[e^{\hat{\beta}\tilde{\phi}}-1]^{2}}
 \ . \label{mel:c}
\end{eqnarray} \pagebreak \\
{\it iv) Contribution of diagram (d)}\\

This diagram involves the 4-gluon vertex, which consists of types of terms
differing in the colour structure: terms involving the structure constants
$f_{ABC}$,
and terms involving the completely symmetric colour couplings $d_{ABC}$. We
denote
the corresponding contributions to $\hat{\Pi}^{(\beta,\mu)}_{44}(\hat{k})$
by  $[\hat{\Pi}^{(\beta,\mu)}_{44}(\hat{k})]_{[f]}$ and
$[\hat{\Pi}^{(\beta,\mu)}_{44}(\hat{k})]_{[d]}$,
respectively.
Consider first $[\hat{\Pi}^{(\beta,\mu)}_{44}(0,\vec{\hat{k}})_{(d)}]_{[f]}$.
After some algebra one finds that
\alpheqn
\begin{equation}
 \left[\hat{\Pi}^{(\beta,\mu)}_{44}(0,\vec{\hat{k}})_{(d)}\right]_{[f]}=
 \frac{3}{4} g^{2}\frac{1}{\hat{\beta}}\sum^{\frac{\hat{\beta}}{2}-1}_{
 \ell=-\frac{\hat{\beta}}{2}}\int^{\pi}_{-\pi}
 \frac{d^{3}\hat{q}}{(2\pi)^{3}}
 \left[ f^{(d)}(e^{i\hat{\omega}^{+}_{\ell}};\hat{q},\hat{k})\right]_{
 [f]}\ , \label{eq:da}
\end{equation}
where
\begin{equation}
 \left[ f^{(d)}(z;\vec{\hat{q}},\vec{\hat{k}})\right]_{[f]}=
 \frac{-c(\vec{\hat{k}})(z^{2}+1)+d(\vec{\hat{k}})(z-1)^{2}+
 P(\vec{\hat{q}},\vec{\hat{k}})z}{
 [z-\bar{z}_{1}][z-\bar{z}_{2}]}
\end{equation}
and
\begin{eqnarray}
 c(\vec{\hat{k}}) &=& 1+2\sum_{j}\cos \hat{k}_{j} \ , \nonumber \\
 d(\vec{\hat{k}}) &=& 1-\frac{1}{3}\sum_{j} \tilde{\hat{k}}_{j}^{2} \ , \\
 P(\vec{\hat{q}},\vec{\hat{k}}) &=& 2+\frac{1}{6}\sum_{j}
 [\widetilde{(\hat{q}+\hat{k})}_{j}^{2} +
 \widetilde{(\hat{q}-\hat{k})}_{j}^{2}] \ . \nonumber
\end{eqnarray}
\reseteqn
Performing the frequency sum in (\ref{eq:da}) one obtains
\begin{eqnarray}
\lefteqn{
 \left[ \hat{\Pi}^{(\beta,\mu)}_{44}(0,\vec{\hat{k}})_{(d)} \right]_{[f]}=
 \frac{3}{4} g^{2} \left\{ -c(\vec{\hat{k}})+
 d(\vec{\hat{k}}) \right. } \nonumber \\
& &\mbox{}+\int^{\pi}_{-\pi}\frac{d^{3}\hat{q}}{(2\pi)^{3}}
 \left. \left[ (c-d)
 \coth\tilde{\phi}+(d-\frac{1}{2}P)\frac{1}{\sinh\tilde{\phi}} \right]
 [1+2\hat{\eta}_{BE}(\tilde{\phi})] \right\} \ .
\end{eqnarray}
Taking the limit $\vec{\hat{k}} \rightarrow 0$ one finds the following
contribution to
the screening mass
\begin{equation}
 (\hat{m}^{2}_{el})^{(d)}_{[f]}=\frac{1}{2}g^{2} \left\{ -9+\frac{1}{
 2}\int^{\pi}_{-\pi}
 \frac{d^{3}\hat{q}}{(2\pi)^{3}} \left[ 17\coth
 \tilde{\phi}+\frac{1}{\sinh
 \tilde{\phi}} \right] [1+2\hat{\eta}_{BE}(\tilde{\phi})] \right\}
 \ . \label{mel:df}
\end{equation}
Next consider the contribution 
$[\hat{\Pi}_{44}^{(\beta,\mu)}(0,\vec{\hat{k}})_{(d)}]_{[d]}$.
It is given by
\alpheqn
\begin{equation}
 \left[\hat{\Pi}^{(\beta,\mu)}_{44}(0,\vec{\hat{k}})_{
 (d)}\right]_{[d]}=-\frac{1}{2}
 g^{2}\frac{1}{\hat{\beta}}\sum^{\frac{\hat{\beta}}{2}-1}_{\ell=-
 \frac{\hat{\beta}}{2}}\int^{\pi}_{-\pi}\frac{d^{3}\hat{q}}{(2\pi)^{3}}
 \left[ f^{(c)}(e^{i\hat{\omega}^{+}_{\ell}};\vec{\hat{q}},\vec{\hat{k}})
 \right]_{[d]}\ ,
\end{equation}
where
\begin{eqnarray}
 \left[ f^{(d)}(z;\vec{\hat{q}},\vec{\hat{k}})\right]_{[d]} &=&\frac{1}{12}
 \left(\frac{20}{3}+d(A)\right) \frac{ K(\vec{\hat{k}})(z-1)^{2}-L(\vec{\hat{
 q}},\vec{\hat{k}})z}{
 [z-\bar{z}_{1}][z-\bar{z}_{2}]}\ , \\
 K(\vec{\hat{k}}) &=& \sum_{j}\tilde{\hat{k}}^{2}_{j}\ , \\
 L(\vec{\hat{q}},\vec{\hat{k}}) &=& \sum_{j}\tilde{\hat{q}}_{j}^{2}
  \tilde{\hat{k}}^{2}_{j}\ ,
\end{eqnarray}
and where $d(A)$ is defined by
\begin{equation}
 \sum^{8}_{E,F=1}[2d_{AFE} d_{BFE}+d_{FFE}d_{ABE}]=d(A)\delta_{AB}\ .
\end{equation}
\reseteqn
Performing the frequency sum one finds
\begin{eqnarray*}
\lefteqn{
 \left[\hat{\Pi}^{(\beta,\mu)}_{44}(0,\vec{\hat{k}}
 )_{(d)}\right]_{[d]} = g^{2}
 \frac{1}{24}\left(\frac{20}{3}+d(A)\right) } \\
& &\times\left\{ -K+\int^{\pi}_{-\pi}
 \frac{d^{3}\hat{q}}{(2\pi)^{3}}\left[ K \coth\tilde{\phi}-
 \frac{ K+\frac{1}{2} L}{\sinh\tilde{\phi}} \right]
 [1+\hat{\eta}_{BE}
 (\tilde{\phi})] \right\}\ .
\end{eqnarray*}
Since $K(\vec{\hat{k}})$ and $L(\vec{\hat{q}},\vec{\hat{k}})$ vanish for
$\vec{\hat{k}} \rightarrow 0$, it does not contribute
to the screening mass, i.e.,
\begin{equation}
 [(\hat{m}_{el})_{(d)}]_{[d]} = 0 \ . \label{mel:dd}
\end{equation}\\
{\it v) Contribution of diagram (e)}\\

The only other diagram possessing a continuum analog is the ghost
loop shown
in fig.\ 1e. Its contribution is given by
\alpheqn
\begin{equation}
\hat{\Pi}^{(\beta,\mu)}_{44}(0,\vec{\hat{k}})_{(e)} = \frac{3}{2}
 g^{2}\frac{1}{\hat{\beta}}
 \sum^{\frac{\hat{\beta}}{2}-1}_{\ell=-
 \frac{\hat{\beta}}{2}}\int^{\pi}_{-\pi}\frac{d^{3}\hat{q}}{(2\pi)^{3}}
 f^{(d)}(e^{i\hat{\omega}^{+}_{\ell}};\vec{\hat{q}},\vec{\hat{k}})\ ,
\end{equation}
where
\begin{equation}
 f^{(e)}(z;\vec{\hat{q}},\vec{\hat{k}}) = -\frac{1}{2}
 \frac{(z^{2}-1)^{2}}{\Pi^{4}_{i=1}[z-\bar{z}_{i}]} \ .
\end{equation}
\reseteqn
Performing the frequency sum one finds that
\alpheqn
\begin{eqnarray}
 \hat{\Pi}^{(\beta,\mu)}_{44}(0,\vec{\hat{k}})_{(e)} &=& \frac{3}{4}g^{2}
 \left\{ -1+\int^{\pi}_{-\pi}
 \frac{d^{3}\hat{q}}{
 (2\pi)^{3}}[g(\tilde{\phi},\tilde{\psi})+g(\tilde{\psi},\tilde{\phi})]
 \right. \nonumber \\
& &\mbox{}+ \left. 4\int^{\pi}_{-\pi}
 \frac{d^{3}\hat{q}}{(2\pi)^{3}}g(\tilde{\phi},\tilde{\psi})
 \hat{\eta}_{BE}(\tilde{\phi}) \right\} \ , \label{eq:ea}
\end{eqnarray}
where
\begin{equation}
 g(\tilde{\phi},\tilde{\psi}) = \frac{\sinh \tilde{\phi}}{\cosh\tilde{\phi} -
 \cosh\tilde{\psi}}\ .
\end{equation}
\reseteqn
This function is again singular for $\vec{\hat{k}} \rightarrow 0$. 
To compute the
limit we
proceed as discussed earlier and write (\ref{eq:ea}) in the form
\begin{eqnarray*}
 \hat{\Pi}^{(\beta,\mu)}_{44}(0,\vec{\hat{k}})_{(e)} &=& \frac{3}{4}g^{2}
 \left\{  -1+\int^{\pi}_{-\pi}
 \frac{d^{3}\hat{q}}{(2\pi)^{3}}[g(\tilde{\phi},\tilde{\psi})+
 g(\tilde{\psi},\tilde{\phi})]
 [1+2\hat{\eta}_{BE}(\tilde{\phi})]\right. \\
& &\mbox{}+ \left. 2\int^{\pi}_{-\pi}
 \frac{d^{3}\hat{q}}{
 (2\pi)^{3}}g(\tilde{\phi},\tilde{\psi})\bigtriangleup
 \hat{\eta}_{BE}(\tilde{\phi},
 \tilde{\psi}) \right\}\ ,
\end{eqnarray*}
where $\bigtriangleup\hat{\eta}_{BE}(\tilde{\phi},\tilde{\psi})$ 
has been defined in (\ref{def:delbe}).
Taking the limit $\vec{\hat{k}} \rightarrow 0$ 
one obtains
\begin{eqnarray}
(\hat{m}^{2}_{el})_{(e)} &=& \frac{3}{4}g^{2}\left\{ -1 +
 \int^{\pi}_{-\pi}
 \frac{d^{3}\hat{q}}{(2\pi)^{3}}[1+2\hat{\eta}_{BE}(\tilde{\phi})]
 \coth\tilde{\phi} \right. \nonumber \\
& &\mbox{} \left. -2\hat{\beta}\int^{\pi}_{-\pi}
 \frac{d^{3}\hat{q}}{
 (2\pi)^{3}}\frac{e^{\hat{\beta}\tilde{\phi}}}{[e^{\hat{\beta}\tilde{\phi}}-1
 ]^{2}} \right\} \ . \label{mel:e}
\end{eqnarray}
Combining the results (\ref{mel:c}), (\ref{mel:df}), (\ref{mel:dd}) 
and (\ref{mel:e}), we therefore
find that
those diagrams possessing
a continuum analog yield the following contribution to the electric screening
mass in the gluonic sector
\begin{eqnarray}
(\hat{m}^{2}_{el})_{(c+d+e)} &=& g^{2}\left\{ -\frac{3}{4}
 +\frac{1}{2}\int_{-\pi}^{\pi}\frac{d^{3}\hat{q}}{(2\pi)^{3}} 
 \left( \coth\tilde{\phi} -
 \frac{1}{\sinh\tilde{\phi}}\right)[1+2\hat{\eta}_{BE}(\tilde{\phi})]
 \right\} \nonumber \\
& &\mbox{}+6g^{2}\hat{\beta}\int_{-\pi}^{\pi}\frac{d^{3}\hat{q}}{(2\pi)^{3}}
 \frac{e^{\hat{\beta}\tilde{\phi}}}{[e^{\hat{\beta}\tilde{\phi}}-1]^{2}}
 \ . \label{mel:cde}
\end{eqnarray}

The computation of the remaining contributions arising from diagrams (f)
and (g),
which are a consequence of the lattice regularization, is
straight forward. One finds that they cancel the first term in
(\ref{mel:cde}).
Hence the gluonic sector (G) contributes as follows to the screening
mass
\begin{equation}
 [\hat{m}_{el}^{2}(\hat{\beta},\hat{\mu},\hat{m})]_{G}=
 6g^{2}\hat{\beta}\int^{\pi}_{-\pi}\frac{d^{3}\hat{q}}{(2\pi)^{3}}
 \frac{e^{\hat{\beta}\tilde{\phi}}}{[e^{\hat{\beta}\tilde{\phi}}-1]^{2}}\ .
\end{equation}
In the continuum limit (\ref{mel:lim}) the corresponding expression for the
(dimensioned)
screening mass squared is given by
\begin{eqnarray*}
 (m^{2}_{el})_{G} &=& \lim_{a\rightarrow 0}6g^{2}\beta\int^{\frac{\pi}{a}}_{
 -\frac{\pi}{a}}
 \frac{d^{3}q}{(2\pi)^{3}}\frac{e^{\frac{1}{a}\beta\tilde{\phi}
 (a\vec{q}\, )}}{
 [e^{\frac{1}{a}\beta\tilde{\phi}(a\vec{q}\, )}-1]^{2}} \\
&=& \frac{3}{\pi^{2}}g^{2}\beta\int^{\infty}_{0} dq\, q^{2} 
 \frac{e^{\beta q}}{[e^{\beta q} -1]^{2}}\ ,
\end{eqnarray*}
where $q = |\vec{q}\, |$.
After a partial integration this expression takes the form
\[ (m^{2}_{el})_{G} = \frac{6}{\pi^{2}}g^{2} T^{2}\int^{\infty}_{0} dx 
 \frac{x}{e^{x}-1}\ . \]
Making use of
\[ \int^{\infty}_{0} dx \frac{x^{\alpha-1}}{e^x-1} =
 \Gamma(\alpha)\zeta(\alpha)\ , \ \ \alpha > 1 \ , \]
where $\Gamma(\alpha)$ is the Euler Gamma function, and $\zeta(\alpha)$ the
Riemann
Zeta-function, we recover the well known result:\ 
$(m^{2}_{el})_{G} = g^{2}T^{2}$.
\section{Lattice Artefacts in the Screening Mass}
\setcounter{equation}{0}
In this section we compare the one loop result for the electric screening mass
on the lattice with the continuum. This will provide us with an estimate of the
magnitude of the lattice artefacts to be expected in numerical simulations.
The numerical data we present is for two mass-degenerate quarks.

On a lattice the temperature can be varied by either keeping the lattice
spacing fixed and varying the number $N_{\tau}$ of temporal lattice sites, or
by varying the lattice spacing (or equivalently the coupling) keeping
$N_{\tau}$ fixed. For fixed lattice spacing the dependence of the screening
mass on the temperature, fermion mass and chemical potential is given by
(see eq. (\ref{mel:lim}))
\begin{equation}
 [m_{el}(T,m,\mu,a)]_{latt} = \frac{1}{a}\hat{m}_{el}(\frac{1}{Ta},\mu a,
 ma)
\end{equation}
\begin{figure}[htb]
\leavevmode
\centering
\epsfxsize10cm \epsffile{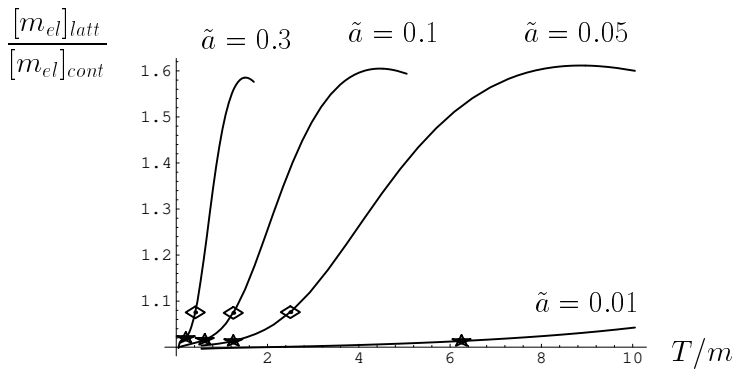} 
\caption{Dependence of $[m_{el}]_{latt}/[m_{el}]_{cont}$ on $\frac{T}{m}$
  for $\mu=0$ and different lattice spacings measured in units of $m^{-1}$.
  Open squares (filled stars) correspond to lattices with $N_{\tau}=8$
  ($N_{\tau}=16$).}
\end{figure}
If the lattice expression is to
approximate the continuum, then the lattice spacing must be small compared
to all
physical length scales in the problem. Hence we must have that
$a \ll \frac{1}{T}$,
$a \ll \frac{1}{m}$ and $a \ll \frac{1}{\mu}$. We therefore expect that
for temperatures $T \ll \frac{1}{a}$ the continuum is well approximated for
$ma \ll 1$ and
$\mu a \ll 1$. 
This is shown in figs. 2 and 3 where we have plotted
$[m_{el}]_{latt}/[m_{el}]_{cont}$ as a function of $T/m$ at $\mu = 0$ and
$\mu/m =1.5$
for various lattice spacings measured
in units of $m^{-1}$. For $\mu = 0$ and $ma \equiv \tilde{a} \in [0, .3]$,
the deviation of this ratio from
unity is seen to be at most
$1.75\%$ for $\frac{T}{m} \leq \frac{1}{16\tilde{a}}$. The rhs of this
inequality is the temperature associated with a lattice with 16 sites
in the temporal direction. 
For $\frac{T}{m} \approx \frac{1}{8\tilde{a}}$ the deviation is already
$7.5\%$.
The endpoint of the curves
for $\tilde{a} = 0.05$, $0.1$ and $0.3$ correspond to the minimal number
of temporal lattice
sites, i.e. $N_{\tau} =2$, and the open squares (filled stars) to
$N_{\tau} =8$ ($N_{\tau} =16$).
\begin{figure}[htb]
\leavevmode
\centering
\epsfxsize10cm \epsffile{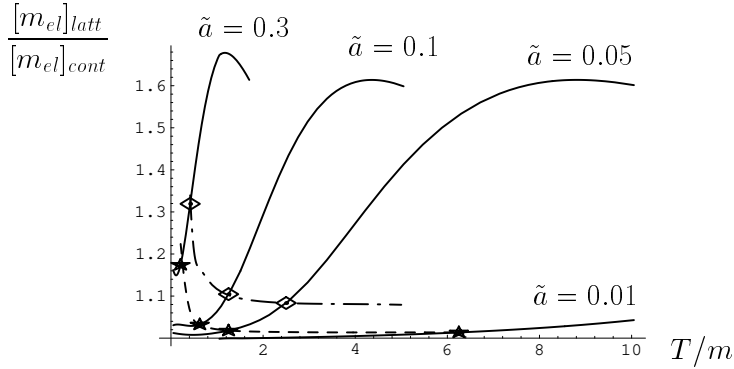}
\caption{Dependence of $[m_{el}]_{latt}/[m_{el}]_{cont}$ on $\frac{T}{m}$
  for $\frac{\mu}{m}=1.5$ and different lattice spacings measured in units 
  of $m^{-1}$.
  Open squares (filled stars) correspond to lattices with $N_{\tau}=8$
  ($N_{\tau}=16$). The dashed-dotted (dashed) lines show the temperature
  dependence for a fixed number of temporal lattice sites, $N_{\tau}=8$
  ($N_{\tau}=16$). }
\end{figure}

For $\mu/m = 1.5$ we must also ensure that $\mu a \ll 1$. We therefore
expect that the allowed range of lattice spacings for achieving an accuracy
of $2\%$ for $T \leq \frac{1}{16 a}$ is now restricted to a smaller interval.
Fig.\ 3 shows that the continuum screening mass is well approximated for
$\tilde{a} < 0.1$ in this temperature range.

In lattice simulations one is interested in determining the electric screening
mass above the deconfining phase transition. It is
extracted from correlators of Polyakov loops [6--9],
or from the long distance behaviour of
the gluon propagator \cite{hell}. 
In these simulations the number of lattice sites
$N_{\tau} \equiv \hat{\beta}$ is
fixed. The electric screening mass in physical units, divided by
the temperature, is then given by
\begin{equation}
 [m_{el}]_{latt}/T = N_{\tau} \hat{m}_{el}(N_{\tau},(\mu/T)N^{-1}_{\tau},
  (m/T)N^{-1}_{\tau})\ ,
\end{equation}
while in the continuum limit ($N_{\tau} \rightarrow \infty$) 
this ratio is just a
function of
$m/T$ and $\mu/T$.
The above conditions for approximating the continuuum now read
(a) $N_{\tau} \gg 1$; (b) $ma = (\frac{m}{T})\frac{1}{N_{\tau}} \ll 1$;
(c) $\mu a = (\frac{\mu}{T})\frac{1}{N_{\tau}} \ll 1$.
We therefore expect that for fixed $N_{\tau}$ the continuum is best
approximated for high temperatures. 
For $\frac{\mu}{m}=1.5$ this is shown in fig.\ 3, where
the temperature dependence of $[m_{el}]_{latt}/[m_{el}]_{cont}$ for
$N_{\tau}=8$ and $N_{\tau}=16$ is given by the dash-dotted and dashed curves
respectively. The strong deviation of this ratio at
low temperatures is due to the fermion loop contribution. 
\begin{figure}[htb]
\leavevmode
\centering
\epsfxsize10cm \epsffile{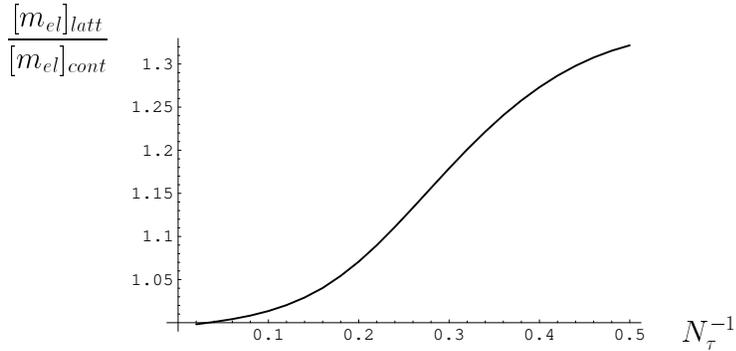} \\
\caption{Dependence of the pure gluonic contribution to
  $[m_{el}]_{latt}/[m_{el}]_{cont}$ on the number of lattice sites $N_{\tau}$.
  The solid line interpolates between different numbers of lattice sites,
  $N_{\tau}$.}
\end{figure}
This is evident
from fig.\ 4, where we have plotted 
$[m_{el}]_{latt}/[m_{el}]_{cont}$ for the pure SU(3) gauge theory for various
values of $N_{\tau}$. This ratio only depends on the number of lattice sites,
$N_{\tau}$. The solid line interpolates between different numbers of
temporal lattice sites.  
The deviation from the continuum is seen to be small already
for $N_{\tau}=8$. For $N_{\tau}=8$, and $N_{\tau}=16$, it is about $2\%$,
and $0.4\%$, respectively.

\begin{figure}[htb]
\leavevmode
\centering
\begin{tabular}{cc}
\epsfxsize7.5cm \epsffile{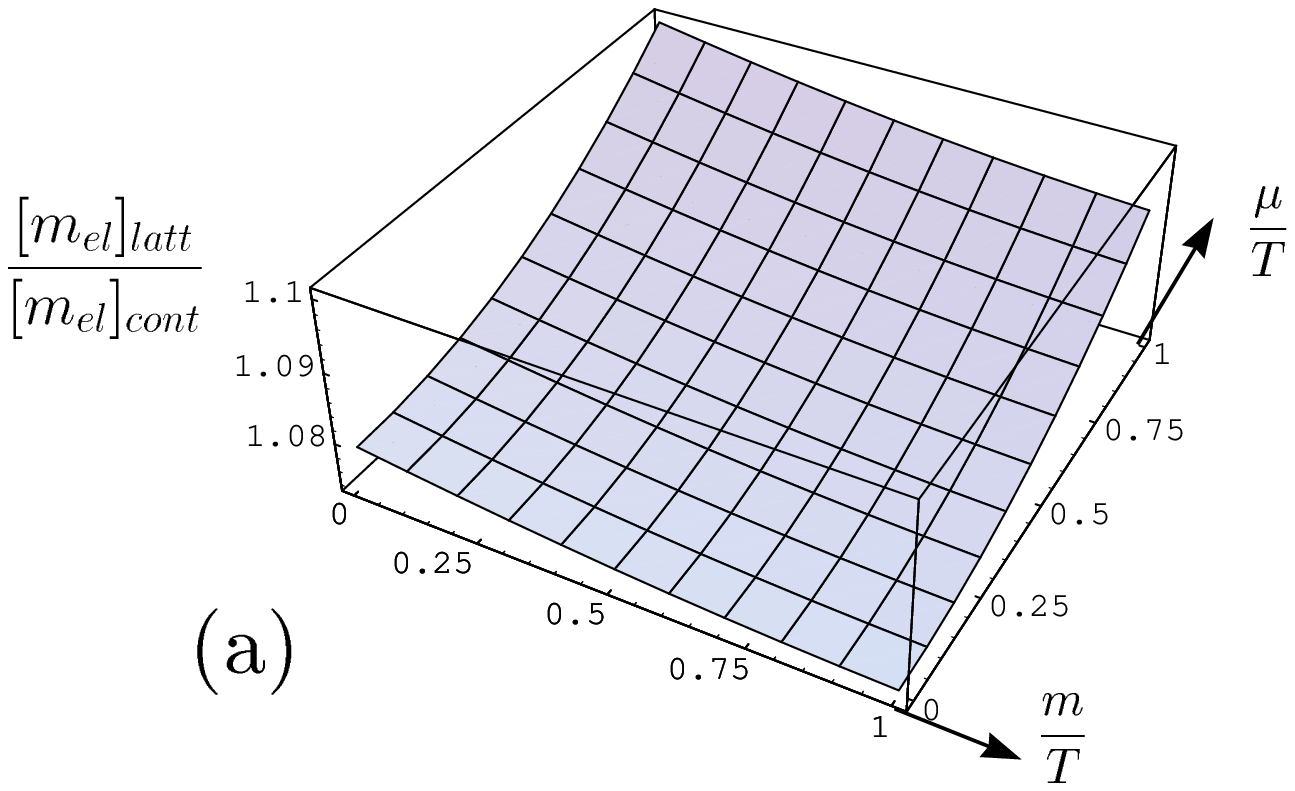} & 
\epsfxsize7.5cm \epsffile{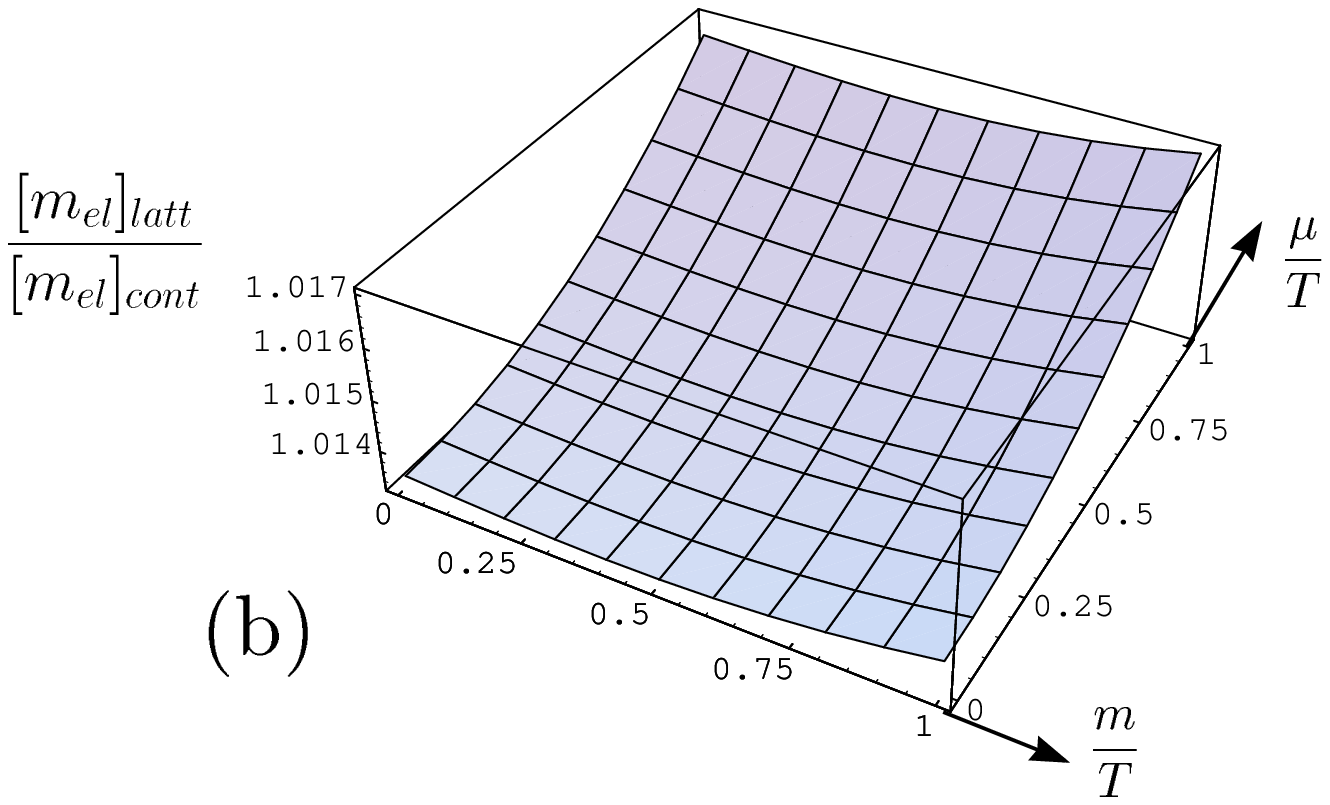} \\
\end{tabular}
\caption{Dependence of $[m_{el}]_{latt}/[m_{el}]_{cont}$ on $\frac{m}{T}$
  and $\frac{\mu}{T}$ for (a) $N_{\tau}=8$, and (b) $N_{\tau}=16$.}
\end{figure}
Finally, in fig.\ 5
we have plotted the ratio $[m_{el}]_{latt}/[m_{el}]_{cont}$ as a function
of $\frac{m}{T}$ and $\frac{\mu}{T}$ for 
$N_{\tau} = 8, 16$. In the parameter
range
considered the above inequalities are well satisfied and the
deviation of this ratio from unity is seen to be at most
$1.7\%$ for $N_{\tau} = 16$,
and $10\%$ for $N_{\tau} = 8$.
The range of values $m/T$ and $\mu/T$ for which the continuum is
well approximated will of course increase with increasing $N_{\tau}$.
\section{Conclusions}
\setcounter{equation}{0}
In this paper we have have computed the electric screening mass for
Wilson fermions in the infinite volume limit for lattice QCD at finite
temperature and chemical potential in one
loop order. The expression we obtained had a very transparent structure in
which the artefacts arising from a finite lattice spacing were concentrated
in two functions which in the naive continuum limit reduced to the
on shell energies of a free quark and gluon. We then studied the
dependence of the lattice
screening mass on the temperature and chemical potential for fixed values of
the lattice spacing. It was found that lattice artefacts
give rise to an enhancement of the screening mass. For $\mu =0$
and lattice spacings $a < .3m^{-1}$ the deviation
of $[m_{el}]_{latt}/[m_{el}]_{cont}$ from the continuum was found to
be less than $1.75\%$ for $\frac{T}{m} \leq \frac{1}{16\tilde{a}}$,
where $\tilde{a}=ma$. For
$\frac{\mu}{m} = 1.5$ a substantially smaller lattice spacing was
required to approximate the continuum. Most of the deviation was found to
be due to the fermion loop contribution.

Since in numerical simulations the temperature dependence of the screening
mass is extracted from a given lattice by varying the coupling
(lattice spacing), we have also studied the dependence of the lattice
screening mass on the temperature and chemical potential for 
fixed $N_{\tau}$. It
was found that for $N_{\tau} =16$, and temperatures larger than the fermion
mass and chemical potential, the continuum screening mass was approximated
to within $1.75\%$. The corresponding
deviation for $N_{\tau} = 8$ was found to be
at most $10\%$ and to be due to the fermion loop contribution.
In the pure SU(3) gauge theory the continuum
was already approximated to $2\%$ for only 8 lattice sites in
the temporal direction.

Our analysis was carried out for infinite lattice volume.
For finite spatial volume the momentum spectrum becomes
discrete and zero momentum modes must be treated separately in a
perturbative expansion. In comparing the data for the electric
screening mass obtained in Monte Carlo simulations with continuum
perturbation theory, finite volume effects need to be
included, while finite lattice spacing effects may, as our analysis
suggests, be negligible
for those couplings and lattice sizes at which the simulations have been
performed.\\ \\
\begin{center} {\bf ACKNOWLEDGEMENTS} \end{center}
We are very grateful to T. Reisz for several discussions and constructive
comments.
\appendix
\appeqn
\appfig
\section{Feynman rules}
\setcounter{equation}{0}
For vanishing temperature and chemical potential the lattice Feynman rules
have been given in \cite{kawa}. 
The expression for the four gluon vertex stated
in that reference is however not quite correct. For completeness sake
we collect in this Appendix the Feynman rules for lattice QCD with Wilson
fermions at finite temperature and chemical potential. In this case
 the fourth component of boson and fermion momenta are replaced by
$k_{4}=\omega_{\ell}^{+}=2\pi \ell/ \beta$ and 
$p_{4}=\omega_{\ell}^{-}+i\mu =(2 \ell+1)\pi / \beta+i\mu$ respectively.
As usual the gluon fields are defined at the middle of the 
links. Following standard conventions, we did not include the
combinatorial factor arising from the symmetrization of the vertices.
The dimensionless
version of these rules, used in section 2, are obtained by setting $a=1$.\\ \\
i) Fermion, gauge field and ghost propagators:\\
\begin{picture}(300,60)
\ArrowLine(20,40)(80,40)
\put(20,40){\circle*{4}}
\put(80,40){\circle*{4}}
\put(80,30){\makebox(0,0)[t]{$a,\alpha$}}
\put(20,30){\makebox(0,0)[t]{$b,\beta$}}
\put(50,30){\makebox(0,0)[t]{$p$}}
\put(90,40){\makebox(0,0)[l]{$\ \ \delta_{ab}\, \frac{\left[ -i/a \, \sum_{\mu} 
 \gam \sin(a p_{\mu})+m(p) \right]_{\alpha\beta}}{1/a^{2} 
 \sum_{\mu} \sin^{2}(a p_{\mu}) +m^{2}(p) }$}}
\end{picture}
\vspace{-3mm} \\
where
\[ m(p) = m+\frac{2r}{a} \sum_{\mu} \sin^{2}(\frac{a}{2}\, 
  p_{\mu}) \]
is the momentum dependent Wilson mass and r the Wilson parameter which we have
set to 1 in the calculation. \\
\begin{picture}(300,60)
\Gluon(20,40)(80,40)54
\put(20,40){\circle*{4}}
\put(80,40){\circle*{4}}
\put(80,30){\makebox(0,0)[t]{$A,\mu$}}
\put(20,30){\makebox(0,0)[t]{$B,\nu$}}
\put(50,30){\makebox(0,0)[t]{$k$}}
\put(90,40){\makebox(0,0)[l]{$\ \ \delta_{AB}\frac{1}{\kt^{2}}\, 
 [ \delta_{\mu\nu}
 - (1-\lambda)\frac{\kt_{\mu}\kt_{\nu}}{\kt^{2}} ] \ \ \ \  ; \ \ \ \
 \kt_{\mu}=\frac{2}{a}\, \sin(\frac{a}{2}\, k_{\mu}) $}}
\end{picture}
\vspace{-3mm} \\
\begin{picture}(300,60)
\DashArrowLine(20,40)(80,40)5
\put(20,40){\circle*{4}}
\put(80,40){\circle*{4}}
\put(80,30){\makebox(0,0)[t]{$A$}}
\put(20,30){\makebox(0,0)[t]{$B$}}
\put(50,30){\makebox(0,0)[t]{$k$}}
\put(90,40){\makebox(0,0)[l]{$\ \ \delta_{AB}\frac{1}{\kt^{2}}$}}
\end{picture} \\
ii) The vertices: \\
\begin{picture}(300,100)
\ArrowLine(60,50)(20,90)
\ArrowLine(20,10)(60,50)
\Gluon(60,50)(110,50)53
\put(85,55){\vector(-1,0){0}}
\put(35,90){\makebox(0,0)[tl]{$\overline{\psi}_{\alpha}^{\, a}(p)$}}
\put(35,10){\makebox(0,0)[bl]{$\psi_{\beta}^{b}(q)$}}
\put(90,60){\makebox(0,0)[bl]{$A_{\mu}^{B}(k)$}}
\put(130,55){\makebox(0,0)[l]{$ -\beta \delta_{0,(-\omega_{\ell_{p}}^{-}+
 \omega_{\ell_{q}}^{-}+\omega_{\ell_{k}}^{+})}\, 
  (2\pi)^{3}\delta_{(p.)}^{3}(-\vp+\vq+\vk\, )$}}
\put(130,35){\makebox(0,0)[l]{$ig (T^{B})_{ab} \{
  (\gam)_{\alpha\beta}\cos(\frac{a}{2}(p+q)_{\mu})-ir\delta_{\alpha
  \beta} \sin(\frac{a}{2}(p+q)_{\mu}) \} $}}
\end{picture} \\
\begin{picture}(300,100)
\ArrowLine(60,50)(20,90)
\ArrowLine(20,10)(60,50)
\Gluon(60,50)(100,90)53
\Gluon(60,50)(100,10)53
\put(80,75){\vector(-1,-1){0}}
\put(80,35){\vector(-1,1){0}}
\put(35,90){\makebox(0,0)[tl]{$\overline{\psi}_{\alpha}^{\, a}(p)$}}
\put(35,10){\makebox(0,0)[bl]{$\psi_{\beta}^{b}(q)$}}
\put(105,90){\makebox(0,0)[tl]{$A_{\mu}^{B}(k)$}}
\put(105,10){\makebox(0,0)[bl]{$A_{\nu}^{C}(r)$}}
\put(130,65){\makebox(0,0)[l]{$-\beta \delta_{0,(-\omega_{\ell_{p}}^{-}+
  \omega_{\ell_{q}}^{-}+\omega_{\ell_{k}}^{+}+\omega_{\ell_{r}}^{+})}\, 
  (2\pi)^{3}\delta_{(p.)}^{3}(-\vp+\vq+\vk+\vr\, ) $}}
\put(130,45){\makebox(0,0)[l]{$ \frac{1}{2}\, ag^{2}
  \delta_{\mu\nu} \{ T^{B}, T^{C} \}_{ab} \{   r\delta_{\alpha\beta}
  \cos(\frac{a}{2}(p+q)_{\mu}) $}}
\put(200,25){\makebox(0,0)[l]{$-i(\gam)_{
  \alpha\beta} \sin(\frac{a}{2}(p+q)_{\mu}) \} $}}
\end{picture} \\
\begin{picture}(300,100)
\DashArrowLine(60,50)(20,90)4
\DashArrowLine(20,10)(60,50)4
\Gluon(60,50)(110,50)53
\put(85,55){\vector(-1,0){0}}
\put(35,90){\makebox(0,0)[tl]{$\overline{c}^{A}(r)$}}
\put(35,10){\makebox(0,0)[bl]{$c^{B}(s)$}}
\put(90,60){\makebox(0,0)[bl]{$A_{\mu}^{C}(k)$}}
\put(130,55){\makebox(0,0)[l]{$ \beta \delta_{0,(-\omega_{\ell_{r}}^{+}+
  \omega_{\ell_{s}}^{+}+\omega_{\ell_{k}}^{+})}\, 
  (2\pi)^{3}\delta_{(p.)}^{3}(-\vr+\vs+\vk\, ) $}}
\put(200,35){\makebox(0,0)[l]{$ ig f_{ABC} \rt_{\mu}
  \cos(\frac{a}{2}\, s_{\mu}) $}}
\end{picture} \\
\begin{picture}(300,100)
\DashArrowLine(60,50)(20,90)4
\DashArrowLine(20,10)(60,50)4
\Gluon(60,50)(100,90)53
\Gluon(60,50)(100,10)53
\put(80,75){\vector(-1,-1){0}}
\put(80,35){\vector(-1,1){0}}
\put(35,90){\makebox(0,0)[tl]{$\overline{c}^{A}(r)$}}
\put(35,10){\makebox(0,0)[bl]{$c^{B}(s)$}}
\put(105,90){\makebox(0,0)[tl]{$A_{\nu}^{D}(q)$}}
\put(105,10){\makebox(0,0)[bl]{$A_{\mu}^{C}(k)$}}
\put(130,60){\makebox(0,0)[l]{$ \beta \delta_{0,(-\omega_{\ell_{r}}^{+}+
  \omega_{\ell_{s}}^{+}+\omega_{\ell_{k}}^{+}+\omega_{\ell_{q}}^{+})}\, 
  (2\pi)^{3}\delta_{(p.)}^{3}(-\vr+\vs+\vk+\vq\, ) $}}
\put(200,40){\makebox(0,0)[l]{$ \frac{1}{12}\, 
  a^{2}g^{2}
  \delta_{\mu\nu} \{ t^{C}, t^{D} \}_{AB} \rt_{\mu}\st_{\mu} $}}
\end{picture} 
\vspace{-7mm} \\
\begin{picture}(300,100)
\Gluon(20,60)(100,60)55
\Line(50,50)(70,70)
\Line(70,50)(50,70)
\put(40,65){\vector(1,0){0}}
\put(80,65){\vector(-1,0){0}}
\put(40,45){\makebox(0,0)[t]{$A_{\mu}^{B}(k)$}}
\put(80,45){\makebox(0,0)[t]{$A_{\nu}^{C}(q)$}}
\put(130,50){\makebox(0,0)[l]{$ -\beta \delta_{0,(\omega_{\ell_{k}}^{+}+
  \omega_{\ell_{q}}^{+})}\, (2\pi)^{3}
  \delta_{(p.)}^{3}(\vk+\vq\, ) \, \frac{1}{4a^{2}}
  \, g^{2} \delta_{BC}
  \delta_{\mu\nu}  $}}
\end{picture}
\vspace{-6mm} \\
\begin{picture}(300,100)
\Gluon(20,90)(60,50)53
\Gluon(20,10)(60,50)53
\Gluon(60,50)(110,50)53
\put(85,55){\vector(-1,0){0}}
\put(40,75){\vector(1,-1){0}}
\put(40,35){\vector(1,1){0}}
\put(40,90){\makebox(0,0)[tl]{$A_{\nu}^{A}(k)$}}
\put(35,10){\makebox(0,0)[bl]{$A_{\mu}^{B}(q)$}}
\put(90,60){\makebox(0,0)[bl]{$A_{\lambda}^{C}(r)$}}
\put(130,70){\makebox(0,0)[l]{$ \beta \delta_{0,(\omega_{\ell_{k}}^{+}+
  \omega_{\ell_{q}}^{+}+\omega_{\ell_{r}}^{+})}\, 
  (2\pi)^{3}\delta_{(p.)}^{3}(\vk+\vq+\vr\, ) $}}
\put(130,50){\makebox(0,0)[l]{$ i g
  f_{ABC}  
  \{ \delta_{\nu\lambda} \widetilde{(k-r)}_{\mu} \cos(\frac{a}{2}\, 
  q_{\nu})  $}}
\put(130,30){\makebox(0,0)[l]{$ +\delta_{\mu\nu} \widetilde{(q-k)}_{\lambda} \cos(\frac{a}{2}\, 
  r_{\mu}) 
  +\delta_{\lambda\mu} \widetilde{(r-q)}_{\nu} \cos(\frac{a}{2}\, 
  k_{\lambda}) \} $}}
\end{picture} \\
\begin{picture}(300,100)
\Gluon(20,90)(60,50)53
\Gluon(20,10)(60,50)53
\Gluon(60,50)(100,90)53
\Gluon(60,50)(100,10)53
\put(80,75){\vector(-1,-1){0}}
\put(80,35){\vector(-1,1){0}}
\put(40,75){\vector(1,-1){0}}
\put(40,35){\vector(1,1){0}}
\put(40,90){\makebox(0,0)[tl]{$A_{\mu}^{A}(k)$}}
\put(35,10){\makebox(0,0)[bl]{$A_{\nu}^{B}(q)$}}
\put(105,90){\makebox(0,0)[tl]{$A_{\rho}^{D}(s)$}}
\put(105,10){\makebox(0,0)[bl]{$A_{\lambda}^{C}(r)$}}
\put(130,60){\makebox(0,0)[l]{$\beta \delta_{0,(\omega_{\ell_{k}}^{+}+
  \omega_{\ell_{q}}^{+}+\omega_{\ell_{r}}^{+}+\omega_{\ell_{s}}^{+})} \, 
  (2\pi)^{3}\delta_{(p.)}^{3}(\vk+\vq+\vr+\vs\, )$}}
\put(230,40){\makebox(0,0)[l]{$ 
  \tilde{\Gamma}_{\mu\nu\lambda\rho}^{(4),ABCD}
  (k,q,r,s) $}}
\end{picture} \\
where
\begin{eqnarray*}
\lefteqn{
 \tilde{\Gamma}_{\mu\nu\lambda\rho}^{(4)ABCD}(k,q,r,s)=
  \frac{1}{12}g^{2} \left\{  \ 
  \left\{ \sum_{E}f_{ABE}f_{CDE} \right. \right.  } \\
 & & \left\{
     \mbox{}-\delta_{\mu\lambda}\delta_{\nu\rho}\{ 12\cos(\frac{1}{2}a
      (k-r)_{\nu}\cos(\frac{1}{2}a(q-s)_{\mu}) 
      -a^{4}\kt_{\nu}\qt_{\mu}\rt_{\nu}\st_{\mu} \} \right. \\
 & & \mbox{}+\delta_{\mu\rho}\delta_{\nu\lambda}\{ 12\cos(\frac{1}{2}a
      (k-s)_{\nu}\cos(\frac{1}{2}a(q-r)_{\mu}) 
      -a^{4}\kt_{\nu}\qt_{\mu}\rt_{\mu}\st_{\nu} \} \\
 & & \mbox{}-\delta_{\nu\lambda}\delta_{\nu\rho}2a^{2}
      \widetilde{(s-r)}_{\mu}\kt_{\nu}\cos(\frac{1}{2}aq_{\mu}) \\
 & & \mbox{}+\delta_{\mu\lambda}\delta_{\mu\rho}2a^{2}
      \widetilde{(s-r)}_{\nu}\qt_{\mu}\cos(\frac{1}{2}ak_{\nu}) \\
 & & \mbox{}-\delta_{\mu\nu}\delta_{\mu\rho}2a^{2}
      \widetilde{(q-k)}_{\lambda}\rt_{\rho}\cos(\frac{1}{2}as_{\lambda}) \\
 & & \mbox{}+\delta_{\mu\nu}\delta_{\mu\lambda}2a^{2}
      \widetilde{(q-k)}_{\rho}\st_{\lambda}\cos(\frac{1}{2}ar_{\rho}) \\
 & & \mbox{}-\left. \delta_{\mu\nu}\delta_{\mu\lambda}\delta_{\mu\rho}a^{2}
      \sum_{\sigma}\widetilde{(q-k)}_{\sigma}
      \widetilde{(s-r)}_{\sigma} \right\} \\
 & & \mbox{+ the cyclic permutations: } \\
 & & \left( \begin{array}{ccc}
             (B,q,\nu) & (C,r,\lambda) & (D,s,\rho) \\
             (C,r,\lambda) & (D,s,\rho) & (B,q,\nu)
            \end{array}  \right) \\
 & & \mbox{ and } \left.
     \left( \begin{array}{ccc}
             (B,q,\nu) & (C,r,\lambda) & (D,s,\rho) \\
             (D,s,\rho) & (B,q,\nu) & (C,r,\lambda)
            \end{array}  \right)  \right\} \\
 & & \mbox{}+a^{4} \left\{ \frac{2}{3}(\delta_{AB}\delta_{CD}+
      \delta_{AC}\delta_{DB}+\delta_{AD}\delta_{BC}) \right.\\
 & & \mbox{}+\left.
     \sum_{E}(d_{ABE}d_{CDE}+d_{ACE}d_{DBE}+d_{ADE}d_{BCE}) \right\} \\
 & & \left\{ \delta_{\mu\nu}\delta_{\mu\lambda}\delta_{\mu\rho}
      \sum_{\sigma}\kt_{\sigma}\qt_{\sigma}\rt_{\sigma}\st_{\sigma}
      +\delta_{\mu\nu}\delta_{\lambda\rho}\kt_{\lambda}
      \qt_{\lambda}\rt_{\mu}\st_{\mu} \right. \\
 & & \mbox{}+\delta_{\mu\lambda}\delta_{\nu\rho}\kt_{\nu}\qt_{\mu}
      \rt_{\nu}\st_{\mu}
      +\delta_{\mu\rho}\delta_{\nu\lambda}\kt_{\nu}\qt_{\mu}
      \rt_{\mu}\st_{\nu} 
      -\delta_{\nu\lambda}\delta_{\nu\rho}\kt_{\nu}\qt_{\mu}
      \rt_{\mu}\st_{\mu} \\
 & & \mbox{}\left. \left. 
     -\delta_{\mu\lambda}\delta_{\mu\rho}\kt_{\nu}\qt_{\mu}
      \rt_{\nu}\st_{\nu} 
      -\delta_{\mu\nu}\delta_{\mu\rho}\kt_{\lambda}\qt_{\lambda}
      \rt_{\mu}\st_{\lambda}
      -\delta_{\mu\nu}\delta_{\mu\lambda}\kt_{\rho}\qt_{\rho}
      \rt_{\rho}\st_{\mu} \right\} \right\} \ .
\end{eqnarray*}
\section{Frequency summation formulae}
\setcounter{equation}{0}
In this Appendix we derive summation formulae over Matsubara frequencies 
useful for performing calculations on a lattice, where 
the frequencies are restricted to the Brillouin zone. We will first 
consider the bosonic case and afterwards turn to the fermionic one.
\subsection{Bosonic frequency sums}
We will prove the following summation formula: \vspace{2mm} \\
\hspace*{5mm} Let $g(z)$ be a meromorphic function of a complex variable $z$ 
that is bounded \linebreak 
\hspace*{5mm} for
$|z| \rightarrow \infty$ and has no singularities on the circle $|z|=1$.
Then
\begin{equation}
 \shmg g(e^{i\whp})= -\sum_{i}\frac{\res_{\zli}\left(\frac{1}{z}g(z)\right)}{
 \zli^{\bh} - 1}  \ , \label{sum:bosg}
\end{equation}
\hspace*{5mm} where $\bh$ is a positive even number, $\whp=2\pi n/ \bh$ and 
$\res_{\zli}\left(\frac{1}{z}g(z)\right)$
denotes \linebreak 
\hspace*{5mm} the residue of $\frac{1}{z}g(z)$ at the pole $\zli$. 
\vspace{2mm} \\
Consider the function
\[
 h(\wh) = \frac{i\bh}{\exp(i\bh\wh)-1} 
\]
which has simple poles at $\wh=\whp$, $n \in Z$, with
residue 1. Because of the above conditions on $g(z)$ there exists an
$\epsilon > 0$ such that $g(e^{i\wh})$ has no singularities for
$\im(\wh) \in [-\epsilon,\epsilon]$. Hence
\begin{equation}
 \shmg g(e^{i\whp}) = \frac{1}{2\pi}\oint_{C}d\wh\, g(e^{i\wh})
  \frac{1}{\exp(i\bh\wh)-1} \label{bosein}
\end{equation} 
where C is the integration contour depicted in fig.\ \ref{fig:bosa}. 
\begin{figure}[ht]
\hspace*{30mm}
\begin{picture}(200,100)
\LongArrow(20,50)(180,50)
\LongArrow(100,10)(100,90)
\DashArrowLine(50,30)(150,30)5
\DashArrowLine(150,70)(50,70)5
\DashCArc(150,50)(20,270,90)2
\DashCArc(50,50)(20,90,270)2
\multiput(50,50)(10,0){10}{\circle*{3}}
\LongArrow(185,60)(185,70)
\LongArrow(185,60)(185,50)
\Line(50,45)(50,55)
\Line(150,45)(150,55)
\put(50,40){\makebox(0,0)[t]{$-\pi$}}
\put(150,40){\makebox(0,0)[t]{$\pi$}}
\put(190,60){\makebox(0,0)[l]{$\epsilon$}}
\put(120,75){\makebox(0,0)[b]{C}}
\end{picture} 
 \caption{Integration contour C in eq. (\ref{bosein})}
\label{fig:bosa}
\end{figure}
Introducing the variable $z=\exp(i\wh)$ we obtain
\begin{equation}
 \shmg g(e^{i\whp})=\frac{1}{2\pi i} \sum_{j=1}^{2}
 \oint_{C_{j}}dz\, \frac{g(z)}{z}
 \frac{1}{z^{\bh}-1} \ , \label{boszwei}
\end{equation}
where $C_{1}$ and $C_{2}$ are the contours shown in fig.\ \ref{fig:bosb} 
with the
integration carried out in the indicated sense. 
\begin{figure}[ht]
\hspace*{40mm}
\begin{picture}(150,150)
\LongArrow(10,75)(140,75)
\LongArrow(75,10)(75,140)
\Line(115,70)(115,80)
\Line(35,70)(35,80)
\DashArrowArcn(75,75)(30,135,315)2
\DashArrowArcn(75,75)(30,315,135)2
\DashArrowArc(75,75)(50,45,225)2
\DashArrowArc(75,75)(50,225,45)2
\put(35,65){\makebox(0,0)[t]{-1}}
\put(115,65){\makebox(0,0)[t]{1}}
\put(100,80){\makebox(0,0)[rb]{$C_{1}$}}
\put(130,80){\makebox(0,0)[lb]{$C_{2}$}}
\end{picture} 
\caption{Integration contours $C_{1}$ and $C_{2}$ in eq. (\ref{boszwei})}
\label{fig:bosb}
\end{figure}
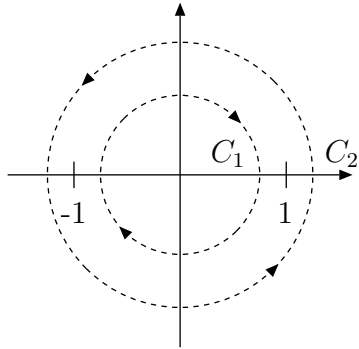

Since $ \lim_{|z| \rightarrow \infty} z^{-\bh}g(z) = 0 $ we can distort the
outer integration contour to infinity taking into account the singularities
of $g(z)$ for $|z|>1$. In the case where $g(z)$ is a meromorphic function
(\ref{sum:bosg}) follows immediately.
\subsection{Fermionic frequency sums}
In the fermionic case we will proceed in an analogous way to prove  
the following summation formula: \vspace{2mm}\\
\hspace*{5mm} Let $g(z)$ be a meromorphic 
function of a complex variable $z$ that is bounded \linebreak
\hspace*{5mm} for 
$|z| \rightarrow \infty$ and has no singularities on the circle 
$|z|=\exp(-\muh)$. Then
\begin{equation}
 \shmmg g(e^{i(\whm+i\muh)})= \sum_{i}\frac{\res_{\zli}\left(\frac{1}{z}g(z)
 \right)}{
  e^{\bh\muh}\zli^{\bh} + 1} \ , \label{sum:fermg}
\end{equation}
\hspace*{5mm} where $\whm=(2m+1)\pi / \bh$. \vspace{2mm} \\
Consider the function
\[
 \tilde{h}(\wh) = \frac{-i\bh}{\exp(i\bh(\wh-i\muh))+1} 
\]
which has simple poles at 
$\wh=\whm+i\muh$, $ m \in Z$, with residue 1. Because of the conditions on 
$g(z)$ there exists an $\epsilon >0$ such that $g(e^{i\wh})$ has no 
singularities for $\im(\wh) \in [\muh-\epsilon,\muh+\epsilon]$. 
Hence
\begin{equation}
 \shmmg g(e^{i(\whm+i\muh)}) = -\, \frac{1}{2\pi}\oint_{C}d\wh\,
  g(e^{i\wh})\frac{1}{\exp(i\bh(
  \wh-i\muh))+1} \label{fermeins}
\end{equation}
where the integration contour C is depicted in fig.\ \ref{fig:fera}.
\begin{figure}[ht]
\hspace*{30mm}
\begin{picture}(200,100)
\LongArrow(20,20)(180,20)
\LongArrow(100,10)(100,90)
\DashArrowLine(50,30)(150,30)5
\DashArrowLine(150,70)(50,70)5
\DashCArc(150,50)(20,270,90)2
\DashCArc(50,50)(20,90,270)2
\multiput(55,50)(10,0){10}{\circle*{3}}
\LongArrow(185,60)(185,70)
\LongArrow(185,60)(185,50)
\Line(50,15)(50,25)
\Line(150,15)(150,25)
\Line(95,50)(105,50)
\put(50,10){\makebox(0,0)[t]{$-\pi$}}
\put(150,10){\makebox(0,0)[t]{$\pi$}}
\put(90,50){\makebox(0,0)[r]{$\mu$}}
\put(190,60){\makebox(0,0)[l]{$\epsilon$}}
\put(120,75){\makebox(0,0)[b]{C}}
\end{picture} 
\caption{Integration contour C in eq. (\ref{fermeins})}
\label{fig:fera}
\end{figure}
Introducing again the variable
 $z=\exp(i\wh)$ we obtain
\[ \shmmg g(e^{i(\whm+i\muh)})=-\, \frac{1}{2\pi i}\sum_{j=1}^{2}
 \oint_{C_{j}}dz\, 
  \frac{g(z)}{z}
 \frac{1}{e^{\bh\muh}z^{\bh} + 1} \ , \]
where the integration contours $C_{1}$ and $C_{2}$ are as depicted in 
fig.\ \ref{fig:bosb}  
except that the radii are changed to $e^{-\muh-\epsilon}$ and
$e^{-\muh+\epsilon}$ respectively.
Hence, for the same reason
as in the bosonic case, we can distort the contour 
$C_{2}$ to infinity and are led, for a meromorphic function $g(z)$, to the
summation formula (\ref{sum:fermg}).


\begin{thebibliography}{99}
\bibitem{kala}
 O.K. Kalashnikov and V.V. Klimov, Yad. Fiz. {\bf 31}, 1357 (1980)
 [Sov. J. Nucl.
 Phys. {\bf 31}, 699 (1980)].
\bibitem{rebh}
 A.K. Rebhan, Phys. Rev. {\bf D48}, R3967 (1993); {\bf D52}, 2994 (1995);
 Nucl.
 Phys. {\bf B430}, 319 (1994).
\bibitem{lind}
 A.D. Linde, Phys. Lett. {\bf 96B}, 289 (1980); see also D. Gross, R.
 Pisarski
 and L. Yaffe, Rev. Mod. Phys. {\bf 53}, 43 (1981).
\bibitem{braa}
 E. Braaten and R.D. Pisarski, Phys. Rev. Lett. {\bf 64}, 1338 (1990);
 Nucl.
 Phys. {\bf B337}, 569 (1990).
\bibitem{kobe}
 R. Kobes, G. Kunstatter and A. Rebhan, Phys. Rev. Lett. {\bf 64},
 2992
 (1990);  Nucl. Phys. {\bf B355}, 1 (1991).
\bibitem{gao}
 M. Gao, Phys. Rev. {\bf D41}, 626 (1990).
\bibitem{irba}
 A. Irb\"{a}k, P. LaCock, D. Miller, B. Petersson and T. Reisz, Nucl. Phys.
 {\bf B363}, 34 (1991).
\bibitem{klae}
 L. K\"{a}rkk\"{a}inen, P. LaCock, B. Peterson and T. Reisz, Nucl. Phys.
 {\bf 395},
 733 (1993).
\bibitem{enge}
 J. Engels, V.K. Mitrjushin and T. Neuhaus, Nucl. Phys. {\bf B440}, 555
 (1995).
\bibitem{hell}
 U.M. Heller, F. Karsch and J. Rank, Phys. Lett. {\bf B355}, 511 (1995).
\bibitem{pete}
 B. Petersson and T. Reisz, Nucl. Phys. {\bf B353}, 757 (1991).
\bibitem{piet} 
 R. Pietig, Master Thesis (Heidelberg, 1994). \nopagebreak
\bibitem{kawa}
 H. Kawai, R. Nakayama and K. Seo, Nucl. Phys. {\bf B189}, 40 (1981).
\end{thebibliography}
\end{document}